\documentstyle[preprint,cite,aps]{revtex}
\newcommand{\be}{\begin{equation}}
\newcommand{\ee}{\end{equation}}
\newcommand{\bea}{\begin{eqnarray}}
\newcommand{\eea}{\end{eqnarray}}
\newcommand{\ba}{\begin{array}}
\newcommand{\ea}{\end{array}}

\newcommand{\bb}{\bibitem}

\begin{document}
\draft
\tightenlines

\title{\bf A new picture of the Lifshitz critical behavior}
\author{Marcelo M. Leite\footnote{e-mail:leite@fis.ita.br}}
\address{{\it Departamento de F\'\i sica, Instituto
Tecnol\'ogico de Aeron\'autica, Centro T\'ecnico Aeroespacial,
12228-900, S\~ao Jos\'e dos Campos, SP, Brazil}}
\maketitle

\vspace{0.2cm}
\begin{abstract}
{\it New field theoretic renormalization group methods are developed to
describe in a unified fashion the critical exponents of an $m$-fold Lifshitz
point at the two-loop order in the anisotropic ($m\neq d$) and isotropic
($m=d$ close to 8) situations. The general theory is illustrated for
the $N$-vector $\phi^{4}$ model describing a $d$-dimensional system.
A new regularization and renormalization procedure is presented for both
types of Lifshitz behavior. The anisotropic cases are formulated with two
independent renormalization group transformations. The description of the
isotropic behavior requires only one type of renormalization group
transformation. We point out the conceptual advantages implicit in this
picture and show how this framework is related to other previous
renormalization group treatments for the Lifshitz problem. The
Feynman diagrams of arbitrary loop-order can be performed analytically
provided these integrals are considered to be homogeneous functions of the
external momenta scales. The anisotropic universality class ($N,d,m$)
reduces easily to the Ising-like ($N,d$) when $m=0$. We show that the
isotropic universality class ($N, m$) when $m$ is close to 8 cannot be
obtained from the anisotropic one in the limit $d \rightarrow m$ near 8.
The exponents for the uniaxial case $d=3$, $N=m=1$ are in good agreement
with recent Monte Carlo simulations for the ANNNI model.}
\end{abstract}

\vspace{1cm}
\pacs{PACS: 75.40.-s; 75.40.Cx; 64.60.Kw}

\newpage
\section{Introduction}

Formulated for the first time in 1975 by Hornreich, Luban and Shtrikman in
the context of magnetic systems\cite{Ho-Lu-Sh,Ho}, the Lifshitz critical
behavior has encountered applications in many real physical systems.
Some examples include high-$T_{c}$ superconductors \cite{3,4,5}, ferroelectric
liquid crystals \cite{Ra, Za, Ska}, uniaxial ferroelectrics \cite{Vy},
some types of polymers \cite{Fre,Ba1, Ne, Ba2} and magnetic materials
\cite{Yo, Yok, Sh, Be}. In
particular, the confluence of a disordered, a uniformly ordered and a
modulated ordered phase characterizes the special critical point associated
to this critical behavior, known as the Lifshitz point. The modulated phase
possesses a fixed equilibrium wavevector which vanishes continuously as the
Lifshitz point is approached. When the components of this wavevector span an
$m$-dimensional subspace, the system under consideration displays an $m$-fold
Lifshitz critical behavior. When the order parameter has $N$ components, and
the space dimension of the system is $d$, the Lifshitz universality class is
characterized by the set $(N,d,m)$ . Whether $m\neq d$, the system presents
the anisotropic Lifshitz critical behavior. Otherwise, the $m=d$ case denotes
the isotropic Lifshitz critical behavior. The isotropic case $m=d$ near 8
can be treated, using similar theoretical tools, along the same lines of the
anisotropic case. Thus, we shall focus our attention in these two types of
critical behavior.

In magnetic systems, the uniaxial $(m=1)$ Lifshitz behavior can be
described by an axially next-nearest-neighbor Ising model (ANNNI)
\cite{Se1, Se2}, which consists of a spin-1/2 Ising model on a cubic
lattice with nearest-neighbor interactions as well as next-nearest-neighbor
antiferromagnetic couplings along one single lattice axis, the competing
axis. The competition between the ferro and antiferromagnetic interactions
in this system provokes a different critical behavior when compared to the
pure Ising-like behavior. The magnetic compound $MnP$ has been studied
extensively in recent years, confirming the appearance of the uniaxial
$(m=1)$ Lifshitz behavior which was obtained from theoretical
\cite{Yo, Yok, Sh} as well as experimental \cite{Be} investigations.

This model can be generalized by allowing the
next-nearest-neighbor antiferromagnetic couplings along $m$ directions, which
represents a typical $m$-axial Lifshitz critical behavior. In case $m \neq d$,
the system naturally admits two independent correlation lengths, namely
$\xi_{L2}$ associated to spatial directions perpendicular to the competing
axes, and $\xi_{L4}$ associated to directions parallel to the
$m$-dimensional competition subspace. At the Lifshitz
point these two correlation lengths become related. In the isotropic
behavior $m=d$ close to 8, there is only one correlation length $\xi_{L4}$.

The field-theoretical representation of this model can be expressed
in terms of a modified $\lambda\phi^{4}$ theory containing  higher
derivative terms along the $m$-competing directions. It is given by
the following bare Lagrangian density \cite{AL1} :
\begin{equation}\label{1}
L = \frac{1}{2}|\bigtriangledown_{m}^{2} \phi_0\,|^{2} +
\frac{1}{2}|\bigtriangledown_{(d-m)} \phi_0\,|^{2} +
\delta_0  \frac{1}{2}|\bigtriangledown_{m} \phi_0\,|^{2}
+ \frac{1}{2} t_{0}\phi_0^{2} + \frac{1}{4!}\lambda_0\phi_0^{4} .
\end{equation}

The field theory  treatment turns out to be simpler at the Lifshitz point,
where $T=T_{L}$ and $\delta_{0}=0$. In particular, the functional integral
representation permits, at the Lifshitz point, the decoupling of the momentum
integrals parallel and perpendicular to the competing axes. It would be
interesting to find out
whether this condition could make it possible the evaluation of Feynman
diagrams
to any desired order in a perturbative expansion. Then, the critical
properties of the system, like critical exponents, amplitude ratios and other
universal amounts could be obtained analytically utilizing the renormalization group analysis along with $\epsilon_{L}$-expansion methods. We shall consider
this problem from a rather new perspective, which allows a solution in
perturbation theory, and which provides an analytic tool that may prove
useful in order to figure out the Lifshitz critical behavior in its
complete generality.

In this work we present a detailed construction of the new renormalization
group description for the anisotropic and isotropic cases. This approach
was inspired by an earlier suggestion made by Wilson \cite{Wilson} in order
to obtain the critical exponents corresponding to correlations parallel or
perpendicular to the competing axes in a manifestly independent manner.
This framework was set forth in a previous letter \cite{Le1}. We discuss
the fundamental issues concerning this new renormalization group (RG)
analysis leading to new scaling relations to the isotropic case, whereas
in the anisotropic case it is shown that these relations are equivalent to
previous scaling laws already derived.

In the anisotropic behaviors, the existence of the correlation lengths
$\xi_{L2}$ and $\xi_{L4}$ induces
two independent characteristic external momenta scales which in turn
are used to fix the renormalized field theory in the infrared regime. Since
the theory is massless in the Lifshitz critical temperature $T_{L}$, the
renormalized vertex parts have to be defined at nonvanishing external momenta
scales. We denote by $\kappa_{1}$ the external momenta scale along the
quadratic (noncompeting) $(d-m)$ directions, whereas $\kappa_{2}$ is the
external momenta scale along the quartic (competing) $m$-dimensional subspace. These external momenta scales originate two independent renormalization
group flows in the parameter space. The renormalized
coupling constants flow to two independent fixed points, depending
whether the renormalization group transformation is over $\kappa_{1}$ or
$\kappa_{2}$. At the loop order considered here they are shown to be the
same, but it is suggested that this feature is preserved when the analysis
is carried out for arbitrary loops. On the other hand, the isotropic case is
characterized solely by the correlation length $\xi_{L4}$. It induces only
one characteristic external momenta scale, denoted here by $\kappa_{3}$
which is used to fix the renormalized vertex parts.

Moreover, we calculate all the
critical exponents at least at $O(\epsilon_{L}^{2})$ using a novel technique of
solving higher-loop Feynman diagrams inspired in this new renormalization
group program. Our analysis is performed entirely in momentum space, which
is particularly suitable to tackle this problem. The Feynman diagrams are
carried out with the help of a new approximation in the quartic
momenta subspace, which is the most general approximation consistent with
homogeneity. We will show how a former two-loop approximation presented
earlier in the calculation of the critical exponents perpendicular to the
competing axes \cite{AL1, AL2} can be understood in terms of
those calculated here using this more elaborate procedure. It is also shown
that the relations among the correlation length exponents parallel and
perpendicular to the competing axes, namely $\nu_{L4}= \frac{1}{2} \nu_{L2}$,
and the anomalous dimensions of the fields, $\eta_{L2}= \frac{1}{2} \eta_{L4}$
are  exact at the loop order considered in the present paper. This confirms
the strong anisotropic scale invariance predicted before for this sort of
system \cite{He}.

In Sec. II we set the formalism by defining the normalization conditions for
the $m$-axial Lifshitz critical behavior for the anisotropic and isotropic
criticalities. We show that two sets of normalization conditions can
naturally describe the anisotropic situation without the need of introducing
other dimensionful constants into the analysis.

Section III contains the discussion of the new renormalization
analysis for the anisotropic case in directions perpendicular to the
competition axes, as well as the new renormalization group
description along directions parallel to the competing subspace.

The renormalization group treatment for the isotropic case is the subject
of Sec. IV. The scaling laws are then obtained for this type of critical
behavior. Since the scaling relations are different from those appearing
in the anisotropic behavior, we point out that the isotropic and anisotropic
behaviors are independent and cannot be obtained from each other.

The evaluation of Feynman integrals is presented in Sec. V. We perform the
one-, two-, and three-loop integrals using two different approximations.
The first approximation introduced in \cite{AL1, AL2} is suitable to perform
two- and three-loop integrals in order to obtain the critical exponents
perpendicular to the competing axes, since it preserves the homogeneity of
the Feynman integrals in the external momenta perpendicular to the competing
axes \cite{AL3}. On the other hand, a new approximation is presented here
which preserves the homogeneity of the Feynman diagrams not only in the
external momenta perpendicular to the competing axes, {\it but also in the
external momenta parallel to the competing $m$-dimensional subspace.} Using
a simple condition in the competing subspace, we calculated these integrals
for arbitrary external momenta.

In Sec. VI we calculate all the critical exponents for the
anisotropic case using the scaling relations derived in Sec. III.
It would be interesting to obtain the critical exponents using more than
one renormalization condition in order to check their correctness. This is
done in this section and in the following one. We also discuss our results
comparing with alternative field theoretic treatments and new Monte Carlo
simulations in $d=3$ in the context of the ANNNI model ($m=1$).

Section VII presents the calculation of all the critical exponents for the
isotropic case utilizing the new scaling relations obtained in Sec. IV.
The analytical expressions obtained are new and to our knowledge are
presented here for the first time.

Finally, the conclusions are presented in Sec. VIII and further
applications of the method described in this work are pointed out.

\section{Normalization conditions for the Lifshitz critical behavior}

From the bare Lagrangian given in (1) we can define renormalized quantities
in terms of bare ones through the use of renormalization constants, or
renormalization functions. Here we shall follow closely the standard
$\lambda \phi^{4}$ field-theoretic approach. The interested reader
should consult, for example, the Amit's book \cite{amit} or the original
work by Br\'ezin, Le Guillou and Zinn-Justin \cite{BLZ}.
These renormalization functions are fixed by the specification of the
renormalization scheme used in order to define the renormalized theory. The
renormalization functions are defined in terms of the renormalized reduced
temperature and order parameter (magnetization in the context of magnetic
systems) as $t = Z_{\phi^2}^{-1} t_0$, $M = Z_{\phi}^{\frac{-1}{2}} \phi_0$
and will depend on Feynman integrals. If the theory is renormalized at the
critical
temperature $(t=0)$, the infrared divergences instruct us to renormalize
the theory in nonvanishing external momenta. Therefore, the renormalization
constants at the critical temperatute $T_{L}$ will depend on the external
momenta scales involved in the renormalization program.

We first consider the anisotropic behaviors. The Feynman integrals depend on
two external momenta scales. We find convenient to define two sets of
normalization conditions appropriate to calculate the critical exponents
associated to correlations either perpendicular to or along the competing
axes \cite{Le1}. These external momenta scales
were defined above to be $\kappa_{1}$ and $\kappa_{2}$, respectively.

In order to make the calculation easier wherever more than one momentum
remains finite, we choose the momenta at a symmetry point (SP).
The normalization conditions which yield the critical exponents associated
to correlations perpendicular to the competition axes are given by first
setting all the external momenta along the competition axes to zero
($\kappa_{2}=0$). Let $p_{i}$ be the external momenta perpendicular to
the competition axes and associated to a
generic 1PI vertex part. Then, the external momenta along the quadratic
directions are chosen as $ p_{i}. p_{j} = \frac{\kappa_{1}^{2}}{4}
(4\delta_{ij} - 1)$. This leads to $(p_{i} + p_{j})^{2} = \kappa_{1}^{2}$ for
$i \neq j$. The momentum scale of the two-point
function is fixed through $p^{2} = \kappa_{1}^{2} = 1$. Thus, we have the
following set of renormalized 1PI vertex parts:

\begin{mathletters}
\begin{eqnarray}
&& \Gamma_{R}^{(2)}(0,g_{1}) = 0, \\
&& \frac{\partial\Gamma_{R}^{(2)}(p, g_{1})}{\partial p^{2}}|_{p^{2}=\kappa_{1}^{2}} = 1, \\
&& \Gamma_{R}^{(4)}(p_{i}, g_{1})|_{SP} = g_{1}  , \\
&& \Gamma_{R}^{(2,1)}(p_{1}, p_{2}, p, g_{1})|_{\bar{SP}} = 1 , \\
&& \Gamma_{R}^{(0,2)}(p, g_{1})|_{p^{2}=\kappa_{1}^{2}} = 0 .
\end{eqnarray}
\end{mathletters}
Recall that the symmetry point implies that the insertion momentum in Eq.(2d)
satisfies $p^{2} = (p_{1} + p_{2})^{2} = \kappa_{1}^{2}$.

The suitable normalization conditions to dealing with exponents along the
competition axes are defined in a similar fashion. Firstly, one
sets all the external momenta perpendicular to the competition axes to zero
($\kappa_{1}=0$). If $k'_{i}$ is the external momenta along the
competition axes associated to a generic 1PI vertex part, the external
momenta along the quartic directions are chosen as
$ k'_{i}. k'_{j} = \frac{\kappa_{2}^{2}}{4} (4\delta_{ij} - 1)$.
This implies that $(k'_{i} + k'_{j})^{2} = \kappa_{2}^{2}$ for $i \neq j$.
The momentum scale of the two-point function is fixed through
$k'^{2} = \kappa_{2}^{2} = 1$. The analogous set of renormalized 1PI vertex
parts is given by:

\begin{mathletters}
\begin{eqnarray}
&& \Gamma_{R}^{(2)}(0,g_{2}) = 0, \\
&& \frac{\partial\Gamma_{R}^{(2)}(k', g_{2})}
{\partial k'^{4}}|_{k'^{4}=\kappa_{2}^{4}}
= 1, \\
&& \Gamma_{R}^{(4)}(k'_{i}, g_{2})|_{SP} = g_{2}  , \\
&& \Gamma_{R}^{(2,1)}(k'_{1}, k'_{2}, k', g_{2})|_{\bar{SP}} = 1 , \\
&& \Gamma_{R}^{(0,2)}(k', g_{2})|_{k'^{4}=\kappa_{2}^{4}} = 0 .
\end{eqnarray}
\end{mathletters}

Note that, in principle, these two systems of normalization conditions
seem to provide two renormalized coupling constants, which arise as a
consequence of the two independent flow in the renormalization momenta scales
$\kappa_{1}$ and $\kappa_{2}$. Apparently the whole description
works with two coupling constants,
namely $g_{1} = u_{1} (\kappa_{1}^{2})^{\frac{\epsilon_{L}}{2}}$
(and $ \lambda_{1} =  u_{01} (\kappa_{1}^{2})^{\frac{\epsilon_{L}}{2}}$)
associated to the flow in the momenta components perpendicular to the
$m$-dimensional axes, as well as
$g_{2} = u_{2} (\kappa_{2}^{4})^{\frac{\epsilon_{L}}{2}}$
(and $ \lambda_{2} =  u_{02} (\kappa_{2}^{4})^{\frac{\epsilon_{L}}{2}}$)
associated to the flow in the momenta components parallel to the
$m$-dimensional axes. Nevertheless, as will be shown, the
situation simplifies at the fixed point: both couplings will flow to
the same fixed point, at two-loop level, indicating that this must be so
in higher-loop calculations. The conceptual advantage is to treat
independently the flow in the momenta along and perpendicular to the
competition axes using these two coupling constants. Whether this can be
done in a consistent manner is a separate problem, to be tackled in Sec. VI.

The normalization conditions for the isotropic case ($m=d$ near 8) can be
defined analogously as those parallel to the competition axes for the
anisotropic case. The symmetry point is chosen as follows.
If $k'_{i}$ is the external momenta along the competition axes, the external
momenta along the
quartic directions are chosen as $ k'_{i}. k'_{j} = \frac{\kappa_{3}^{2}}{4}
(4\delta_{ij} - 1)$. This implies that
$(k'_{i} + k'_{j})^{2} = \kappa_{3}^{2}$ for $i \neq j$.
The momentum scale of the two-point function is fixed through
$k'^{4} = \kappa_{3}^{4} = 1$. Then we have the following conditions:

\begin{mathletters}
\begin{eqnarray}
&& \Gamma_{R}^{(2)}(0,g_{3}) = 0, \\
&& \frac{\partial\Gamma_{R}^{(2)}(k', g_{3})}
{\partial k'^{4}}|_{k'^{4}=\kappa_{3}^{4}}
= 1, \\
&& \Gamma_{R}^{(4)}(k'_{i}, g_{3})|_{SP} = g_{3}  , \\
&& \Gamma_{R}^{(2,1)}(k'_{1}, k'_{2}, k', g_{3})|_{\bar{SP}} = 1 , \\
&& \Gamma_{R}^{(0,2)}(k', g_{3})|_{k'^{4}=\kappa_{3}^{4}} = 0 .
\end{eqnarray}
\end{mathletters}

Notice that we have not mentioned the quadratic momenta
scale $\kappa_{1}$ in the discussion of the isotropic behavior, for it is
absent in this situation due to the Lifshitz condition $\delta_{0}=0$.

We can write all the renormalization functions and bare coupling constants
in terms of the dimensionless couplings.
Let the label  $\tau = 1,2,3$ refer to the different external momenta scales
involved in the general Lifshitz critical behavior, as discussed above for
different normalization conditions in the anisotropic and isotropic cases.
By expanding the dimensionless bare coupling
constants $u_{o \tau}$ and the renormalization functions $Z_{\phi (\tau)}$,
$\bar{Z}_{\phi^{2} (\tau)} = Z_{\phi (\tau)} Z_{\phi^{2} (\tau)}$ in terms of
the dimensionless renormalized couplings $u_{\tau}$ up to two-loop order as
\begin{mathletters}
\begin{eqnarray}
&& u_{o \tau} = u_{\tau} (1 + a_{1 \tau} u_{\tau} + a_{2 \tau} u_{\tau}^{2}) ,\\
&& Z_{\phi (\tau)} = 1 + b_{2 \tau} u_{\tau}^{2} + b_{3 \tau} u_{\tau}^{3} ,\\
&& \bar{Z}_{\phi^{2} (\tau)} = 1 + c_{1 \tau} u_{\tau} + c_{2 \tau} u_{\tau}^{2} ,
\end{eqnarray}
\end{mathletters}
along with dimensional regularization will be sufficient to determine all the
critical exponents.

\section{Renormalization group analysis for the anisotropic case}

Given one bare theory, described by the Lagrangian (1), different versions of
the renormalized vertices can be constructed out of the original bare vertex
parts. We shall explore now the freedom left in the definition of the
renormalization momenta scales $\kappa_{1}$ and $\kappa_{2}$ in the critical
theory explained in the last section for the anisotropic case.

We start by considering the renormalization group analysis along directions
perpendicular to the competing axes. The renormalized theory is defined with
only one quadratic nonvanishing external momenta scale $\kappa_{1}$.
Let $\Lambda_{1}$ be the associated cutoff corresponding to this subspace.
The renormalized vertex parts for this case are defined in terms of the
normalization constants and the bare vertices as :
\begin{eqnarray}
\Gamma_{R(\tau)}^{(N,L)} (p_{i (\tau)}, Q_{i(\tau)}, g_{\tau}, \kappa_{\tau})
&=& Z_{\phi (\tau)}^{\frac{N}{2}} Z_{\phi^{2} (\tau)}^{L}
(\Gamma^{(N,L)} (p_{i (\tau)}, Q_{i (\tau)}, \lambda_{\tau}, \Lambda_{\tau})\\ \nonumber
&& - \delta_{N,0} \delta_{L,2}
\Gamma^{(0,2)}_{(\tau)} (Q_{(\tau)}, Q_{(\tau)}, \lambda_{\tau}, \Lambda_{\tau})|_{Q^{2}_{(\tau)} = \kappa_{\tau}^{2}})
\end{eqnarray}
where $p_{i (\tau)}$ ($i=1,...,N$) are the external momenta associated to
the vertex functions $\Gamma_{R(\tau)}^{(N,L)}$ with $N$ external legs and
$Q_{i (\tau)}$ ($i=1,...,L$) are the external momenta associated to the
$L$ insertions of $\phi^{2}$ operators. We emphasize that
$p_{i (1)}$ ($i=1,...,N$) refers to the external momenta components
along the $(d-m)$ dimensional subspace perpendicular to the competition axes,
whereas $p_{i (2)}$ are the external
momenta components along the $m$-dimensional competing subspace. From our
normalization conditions, it should be kept in mind that all quantities
presenting a subscript $\tau=1 (2)$ are calculated at zero external momenta
components parallel (perpendicular) to the competing axes and are
characterized by the momenta scale $\kappa_{1} (\kappa_{2})$.
From the last section, $u_{0 \tau}$, $Z_{\phi (\tau)}$ and
$Z_{\phi^{2} (\tau)}$ are represented as power series in $u_{\tau}$.
The renormalization group invariance
of the bare vertex with the momenta scale $\kappa_{\tau}$ implies that :
\begin{equation}
(\kappa_{\tau} \frac{\partial}{\partial \kappa_{\tau}})_{\lambda_{\tau},
\Lambda_{\tau}}
[Z_{\phi (\tau)}^{-\frac{N}{2}} Z_{\phi^{2} (\tau)}^{-L}
(\Gamma_{R(\tau)}^{(N,L)} - \delta_{N,0} \delta _{L,2} \Gamma_{(\tau)}^{(N,L)})] = 0.
\end{equation}
This in turn  yields the following RG equations:
\begin{eqnarray}
&&(\kappa_{\tau} \frac{\partial}{\partial \kappa_{\tau}} +
\bar{\beta_{\tau}}\frac{\partial}{\partial g_{\tau}}
- \frac{1}{2} N \gamma_{\phi (\tau)}(g_{\tau}, \kappa_{\tau}) + \nonumber \\
&& L \gamma_{\phi^{2} (\tau)}(g_{\tau},\kappa_{\tau}))
\Gamma_{R(\tau)}^{(N,L)} (p_{i (\tau)}, Q_{i (\tau)}, \lambda_{\tau},
\Lambda_{\tau}) = \delta_{N,0} \delta_{L,2} (\kappa_{\tau}^{-2 \tau})^{\frac{\epsilon_{L}}{2}} B_{\tau},
\end{eqnarray}
where $B_{\tau}$ is a constant used to
renormalize $\Gamma_{R (\tau)}^{(0,2)}$ and
\begin{mathletters}
\begin{eqnarray}
&& \bar{\beta_{\tau}}(g_{\tau}, \kappa_{\tau}) = (\kappa_{\tau}
\frac{\partial g_{\tau}}{\partial \kappa_{\tau}})_{\lambda_{\tau},
\Lambda_{\tau}}\\
&& \gamma_{\phi (\tau)}(g_{\tau}, \kappa_{\tau}) = (\kappa_{\tau}
\frac{\partial lnZ_{\phi (\tau)}}{\partial \kappa_{\tau}})_{\lambda_{\tau},
\Lambda_{\tau}}\\
&& \gamma_{\phi^{2} (\tau)}(g_{\tau}, \kappa_{\tau}) = (\kappa_{\tau}
\frac{\partial lnZ_{\phi^{2} (\tau)}}{\partial \kappa_{\tau}})_{\lambda_{\tau}, \Lambda_{\tau}}.
\end{eqnarray}
\end{mathletters}
are functions of $g_{\tau}$ and $\kappa_{\tau}$ only,
though they are functions of $\Lambda_{\tau}$ implicitly. Notice that
$\Gamma_{R}^{(0,2)}$ is different from all other vertices since the RG equation
presents an inhomogeneous term in the left hand side due to its additive
renormalization. We shall treat this
additively renormalized vertex part later on. The above expressions
correspond to the limit $\Lambda_{\tau}\rightarrow \infty$, which are
naturally finite, even if $\lambda_{\tau}, Z_{\phi (\tau)}$ and
$Z_{\phi^{2} (\tau)}$ diverge at this limit. It is worth expressing
all of these quantities in terms of dimensionless bare and renormalized
coupling constants. We now turn our attention to discuss the central issue
of the new dimensional considerations which will be useful for the
subsequent dimensional analysis.

Consider the volume element in momentum space for calculating an arbitrary
Feynman integral. It is given by $d^{d-m}q d^{m}k$, where $\vec{q}$
represents a $(d-m)$-dimensional vector perpendicular to the competing
axes and $\vec{k}$ denotes an $m$-dimensional vector along the competing
subspace, respectively. The Lifshitz condition $\delta_{0}=0$ suppresses
the quadratic part of the momentum along the competition axes. Nevertheless,
there is still a contribution from the quartic momenta contained in the
inverse critical $(t=0)$ free propagator
$G_{0}^{(2) -1}(q,k) = (k^{2})^{2} + q^{2}$. In
order to be dimensionally consistent, the canonical dimension in mass units of
both terms in the propagator should be equal. There are two ways out of this
outstanding situation.

The former idea, inspired in Ref.\cite{Ho-Lu-Sh}, is to introduce a
dimensionful constant $\sigma$ in front of the first term in the Lagrangian
(1), along with its renormalization function $Z_{\sigma}$, as was done by
Mergulh\~ao and Carneiro \cite{Mergulho1}. This idea implies that 
the momenta scales parallel or perpendicular to the
competition directions play the same role in this discussion and there
is only one coupling constant. Denoting the components of the quartic 
external momenta with a subscript $\alpha$ and the quadratic
components with a subscript $\beta$, they set the following
renormalization conditions:
\begin{mathletters}
\begin{eqnarray}
&& \frac{\partial\Gamma_{R}^{(2)}(k, -k,\sigma, g, \kappa)}
{\partial k_{\beta}^{2}}|_{k_{\beta}^{2}=\kappa^{2}}=1 \nonumber\\
&& \frac{\partial\Gamma_{R}^{(2)}(k,-k, \sigma, g, \kappa)}
{\partial k_{\alpha}^{4}}|_{\sigma k_{\alpha}^{4}=\kappa^{2}}
= \sigma, \nonumber\\
&& \Gamma_{R}^{(4)}(k_{i}, \sigma, g, \kappa)|_{SP_{\alpha}} = g  , \nonumber\\
&& \Gamma_{R}^{(2,1)}(k_{1}, k_{2}, p, \sigma, g, \kappa)|_{\bar{SP_{\alpha}}} = 1 , \nonumber\\
&& \Gamma_{R}^{(0,2)}(p,-p, \sigma, g, \kappa)|_{\sigma 
p_{\alpha}^{4}=\kappa^{2}} =0  ,\nonumber
\end{eqnarray}
\end{mathletters}
where the $SP_{\alpha}$ means 
$\sigma^{\frac{1}{2}}k_{i \alpha}k_{j \alpha}=
\kappa\frac{(4\delta_{ij} -1)}
{4}$ and was chosen at zero quadratic external momenta. This choice of
renormalization points makes the renormalization constants $\sigma$ 
independent as claimed by those authors \cite{Mergulho2}. However,
$\sigma$ is still a relevant length (momentum) scale and this fact
should be reflected on its dependence in some normalization
constants. Therefore, starting with a dimensionful $\sigma$ 
parameter and making it dimensionless in the end of the calculation as
they chose does not seem to be consistent, since the quartic and 
quadratic momenta scale play the same role and have the same canonical
dimensions in this approach. Notice that the last 4 of these equations
together with the critical theory condition on the renormalized
two-point vertex part naturally defines an independent set of
normalization conditions along the competing axes. In fact, if the
quartic momenta is redefined through 
$\sigma k_{\alpha}^{4} \equiv k_{\alpha}^{' 4}$ such that $\sigma$ is
absorbed in the new quartic momenta, this implies that 
$\sigma^{\frac{1}{2}}k_{i \alpha}k_{j \alpha}=k'_{i \alpha}k'_{j \alpha}=
\kappa'^{2}\frac{(4\delta_{ij} -1)}
{4}$ with $\kappa'\neq \kappa$. Then, one has 5 normalization 
conditions along the quartic subspace as described in the last
section. 

On the other hand, the first of these equations is calculated
at $\sigma k_\alpha^{4} = 0$. Intuitively it should be complemented
with 4 more normalization conditions with nonvanishing external
quadratic momenta perpendicular to the competing subspace. This is
what was done in the last section for directions perpendicular to the 
competing axes. Thus, if we have two different 
momenta scales $\kappa$ and $\kappa'$ and setting them equal is
equivalent to have Mergulh\~ao and Carneiro's renormalization
conditions, with 5 more normalization conditions along the quadratic 
directions. Thus if one trades $\sigma$ by an independent external
quartic momenta scale $\kappa'$, it still remains the 5 extra 
normalization conditions which in their approach are undefined. 
Nevertheless, they recovered the former anisotropic scaling 
relations \cite{Ho-Lu-Sh} using this reasoning. They used their 
symmetry point in order to treat the cases $m=2,6$ in the context of
an $\epsilon_{L}$-expansion \cite{Mergulho2}.

There is an alternative based in a recently proposed
method which does not use the dimensional constant $\sigma$ but
allows the realization of a dimensional redefinition of the momenta
components  along the quartic competing subspace \cite{Le1}. This
later view inspires the subsequent discussion and shall be used throughout
this paper. Let $[\vec{q}] = M$ be the mass dimension of the quadratic
momenta. The consistency of the Lagrangian density (1) on dimensional grounds
requires that $[\vec{k}] = M^{\frac{1}{2}}$. This is equivalent
to perform a dimensional redefinition of the momenta along the competing
axes, as long as the condition $\delta_{0}=0$ is satisfied. The volume
element in momentum space $d^{d-m}q d^{m}k$ has
mass dimension  $[d^{d-m}q d^{m}k]= M^{d-\frac{m}{2}}$. The dimension of the
field is obtained by requiring that the volume integral of the Lagrangian
density (1) is dimensionless in mass units. It follows that
$[\phi] = M^{\frac{1}{2}(d-\frac{m}{2})-1}$.

The $N$-point Green function can be expressed dimensionally as
$[G^{(N)}(x_{1},..., x_{N})] = [\phi]^{N} =
M^{\frac{N}{2}(d-\frac{m}{2}) - N}$. The associated one particle
irreducible (1PI) vertex parts have
dimension in mass units
$[\Gamma^{(N)}(x_{i})] =  [G^{(N)}(x_{i})] [V]^{-N}[G^{(2)}(x_{i})]^{-N} =
M^{\frac{N}{2}(d-\frac{m}{2}) + N}$. In momentum space, the Fourier transform
is obtained by integrating over each one of the coordinates. Removing the
momentum conserving $\delta$-function, we have
$[\Gamma^{(N)}(k_{i})]= M^{N + (d - \frac{m}{2}) -
\frac{N (d - \frac{m}{2})}{2}}$.

As usual, the exponent of $M$ in the above
relations is called the canonical dimension of the quantity. If the physical
quantity $O$ has canonical dimension $[O]=M^{\Delta}$, then under a
transformation of the length scale associated to $M \rightarrow \alpha M$
it implies that $O = \alpha^{\Delta} O$. Therefore, all dimensionfull
parameters are transformed under a transformation in the lengths
(or external momenta). Hence, it is useful to describe the theory in
terms of dimensionless parameters. As the
coupling constants are associated to $\Gamma^{(4)}$, we can write
$g_{\tau} = u_{\tau} (\kappa_{\tau}^{2 \tau})^{\frac{\epsilon_{L}}{2}}$,
and $ \lambda_{\tau} =  u_{0 \tau}
(\kappa_{\tau}^{2 \tau})^{\frac{\epsilon_{L}}{2}}$,
where $\epsilon_{L}= 4 + \frac{m}{2} - d$.

In terms of the dimensionless couplings defined
above, the renormalization group equation can be rewritten as:
\begin{equation}
(\kappa_{\tau} \frac{\partial}{\partial \kappa_{\tau}} +
\beta_{\tau}\frac{\partial}{\partial u_{\tau}}
- \frac{1}{2} N \gamma_{\phi (\tau)}(u_{\tau}) + L \gamma_{\phi^{2} (\tau)}(u_{\tau}))
\Gamma_{R(\tau)}^{(N,L)} = \delta_{N,0} \delta_{L,2} (\kappa_{\tau}^{-2 \tau})^\frac{\epsilon_{L}}{2} B_{\tau}(u_{\tau}) ,
\end{equation}
and from now on we can forget about the cutoffs $\Lambda_{\tau}$,
bearing in mind, however, that they should be kept fixed in all stages of the
analysis. The functions
\begin{mathletters}
\begin{eqnarray}
&& \beta_{\tau} = (\kappa_{\tau}\frac{\partial u_{\tau}}{\partial \kappa_{\tau}}), \\
&& \gamma_{\phi (\tau)}(u_{\tau})  = \beta_{\tau}
\frac{\partial ln Z_{\phi (\tau)}}{\partial u_{\tau}}\\
&& \gamma_{\phi^{2} (\tau)}(u_{\tau}) = - \beta_{\tau}
\frac{\partial ln Z_{\phi^{2} (\tau)}}{\partial u_{\tau}}
\end{eqnarray}
\end{mathletters}
are calculated at fixed bare coupling $\lambda_{\tau}$.
The $\beta_{\tau}$-functions can be cast in a more useful
form in terms of dimensionless quantities, namely,

\begin{equation}
\beta_{\tau} = - \tau \epsilon_{L}(\frac{\partial ln u_{0 \tau}}{\partial u_{\tau}})^{-1}.
\end{equation}
Note that the beta function corresponding to the flow in $\kappa_{2}$ has a
factor of 2 compared to that associated to the flow in $\kappa_{1}$.
As usual, they are power series in $u_{\tau}$, with coefficients which depend
on $\epsilon_{L}$. Let us analyse the simplest case $L=0$. The solution can be
expressed in terms of characteristics. The characteristic equation is given by two independent flows in the coupling constants induced by the flows in the
momenta scale $\kappa_{1}$ and $\kappa_{2}$, i.e,

\begin{equation}
\rho_{\tau} \frac{du_{\tau}(\rho_{\tau})}{d\rho_{\tau}}= \beta(u_{\tau}
(\rho_{\tau})),
\end{equation}
with the initial condition $u_{\tau}(\rho_{\tau}=1) = u_{\tau}$. Using the
characteristic equation for $u_{\tau}$ we can change variables from a variable
$x_{\tau}$ to $u_{\tau}$, through the relation:
\begin{equation}
\int_{1}^{\rho_{\tau}} f(u_{\tau}(x_{\tau})) \frac{d x_{\tau}}{x_{\tau}}=
\int_{u_{\tau}}^{u_{\tau}(\rho_{\tau})} \frac{f(u_{\tau})}{\beta_{\tau}(u_{\tau})}.
\end{equation}
Thus, small values of $x_{\tau}$ in the left-hand-side correspond to the
neighborhood of the zero's of $\beta_{\tau}$ in the right-hand-side.
For the anisotropic case, the solution to the renormalization group equation
reflects the two-parameters group of invariance, and can be expressed in the
form
\begin{equation}
\Gamma_{R (\tau)}^{(N)} (k_{i (\tau)}, u_{\tau}, \kappa_{\tau}) =
exp[-\frac{N}{2} \int_{1}^{\rho_{\tau}} \gamma_{\phi (\tau)}(u_{\tau}(\rho_{\tau}))
\frac{{d x_{\tau}}}{x_{\tau}}]
\;\Gamma_{R (\tau)}^{(N)} (k_{i (\tau)}, u_{\tau}(\rho_{\tau}),
\kappa_{\tau} \rho_{\tau}).
\end{equation}

From our dimensional analysis, the dimensional redefinition of the momenta
along the competing axes results in  an effective space dimension for the
anisotropic case, i.e., $(d-\frac{m}{2})$. Thus, we find
the following behavior for the 1PI vertex parts $\Gamma_{R (\tau)}^{(N)}$
under flows in the external momenta:
\begin{eqnarray}
\Gamma_{R (\tau)}^{(N)} (\rho_{\tau} k_{i (\tau)}, u_{\tau}, \kappa_{\tau})&=&
\rho_{\tau}^{\tau(N + (d-\frac{m}{2}) - \frac{N(d-\frac{m}{2})}{2})}
exp[-\frac{N}{2} \int_{1}^{\rho_{\tau}} \gamma_{\phi (\tau)}(u_{\tau}(\rho_{\tau}))
\frac{{d x_{\tau}}}{x_{\tau}}]\\
&&\Gamma_{R (\tau)}^{(N)} (k_{i (\tau)}, u_{\tau}(\rho_{\tau}),
\kappa_{\tau} \rho_{\tau})\nonumber.
\end{eqnarray}

It is helpful to present the explicit expressions for vertex parts
calculated either at zero quartic external momenta or at vanishing quadratic
external momenta. The renormalized vertex
parts calculated at zero quartic external momenta is given by
\begin{eqnarray}
\Gamma_{R (1)}^{(N)} (\rho_{1} k_{i (1)}, u_{1}, \kappa_{1})&=&
\rho_{1}^{N + (d-\frac{m}{2}) - \frac{N(d-\frac{m}{2})}{2}}
exp[-\frac{N}{2} \int_{1}^{\rho_{1}} \gamma_{\phi (1)}(u_{1}(\rho_{1}))
\frac{{d x_{1}}}{x_{1}}]\\
&&\Gamma_{R (1)}^{(N)} (k_{i (1)}, u_{1}(\rho_{1}),
\kappa_{1} \rho_{1})\nonumber.
\end{eqnarray}
The dependence of the renormalized vertex parts is quadratic in
the external momenta perpendicular to the competing axes. Therefore,
the analysis is completely similar to the pure $\lambda \phi^{4}$-theory,
with the replacement $\epsilon_{L} \rightarrow \epsilon$. From this analysis,
we can identify the label $\tau=1$ with the subscript $L2$. Then, we could
have written $\gamma_{\phi (1)} \equiv \gamma_{\phi (L2)}$ and
$\gamma_{\phi^{2} (1)} \equiv \gamma_{\phi^{2} (L2)}$.

On the other hand, the renormalized vertex parts at
zero quadratic external momenta can be expressed as
\begin{eqnarray}
\Gamma_{R (2)}^{(N)} (\rho_{2} k_{i (2)}, u_{2}, \kappa_{2})&=&
\rho_{2}^{2(N + (d-\frac{m}{2}) - \frac{N(d-\frac{m}{2})}{2})}
exp[-\frac{N}{2} \int_{1}^{\rho_{2}} \gamma_{\phi (2)}(u_{2}(\rho_{2}))
\frac{{d x_{2}}}{x_{2}}]\\
&&\Gamma_{R (2)}^{(N)} (k_{i (2)}, u_{2}(\rho_{2}),
\kappa_{2} \rho_{2})\nonumber.
\end{eqnarray}
The difference is that the canonical dimension is twice as big as
the canonical dimension of the vertex parts calculated at zero quartic
momenta. Then, we can make the identifications
$\gamma_{\phi (2)} \equiv \gamma_{\phi (L4)}$ and
$\gamma_{\phi^{2} (2)} \equiv \gamma_{\phi^{2} (L4)}$.
The last equations imply that a change in the external momenta scale is
equivalent to the multiplication of the vertex function by that scale to the
power of the canonical dimension of the function, followed by a modified
coupling constant, which flows with the characteristic equation, and an
additional factor.

It is interesting to analyse the vertex functions at the infrared
fixed points, since this will determine the scaling laws and the critical
exponents associated to correlations perpendicular and parallel to the
$m$-dimensional competing subspace. The analysis can be carried out by
assuming that there are two independent fixed points, defined
by $\beta_{\tau}(u_{\tau}^{*}) = 0$. The renormalization group equation
leads to a simple scaling property at the fixed points. It implies the
following solution to the vertex functions:
\begin{eqnarray}
\Gamma_{R (\tau)}^{(N)} (\rho_{\tau} k_{i (\tau)}, u_{\tau}^{*}, \kappa_{\tau})&=&
\rho_{\tau}^{\tau (N + (d-\frac{m}{2}) - \frac{N(d-\frac{m}{2})}{2}) -\frac{N}{2} \gamma_{\phi (\tau)}(u_{\tau}^{*})}\\
&&\Gamma_{R (\tau)}^{(N)} (k_{i (\tau)}, u_{\tau}^{*},\kappa_{\tau})\nonumber.
\end{eqnarray}

For $N=2$, we have
\begin{equation}
\Gamma_{R (\tau)}^{(2)} (\rho_{\tau} k_{(\tau)}, u_{\tau}^{*}, \kappa_{\tau}) =
\rho_{\tau}^{2 \tau - \gamma_{\phi (\tau)}(u_{\tau}^{*})}\Gamma_{R (\tau)}^{(2)} (k_{(\tau)},u_{\tau}^{*}, \kappa_{\tau}).
\end{equation}
The quantity $\gamma_{\phi (\tau)}(u_{\tau}^{*})$ can be interpreted in the
following way. If the field theory is free, a change in the external
momenta scale will produce a change in the free vertex  $\Gamma_{(\tau)}^{(N) 0}$ which scales with the canonical dimension of the vertex, that is
\begin{equation}
\Gamma_{(\tau)}^{(N) 0}(\rho_{\tau}k_{i (\tau)}) = \rho_{\tau}^{\tau (N + (d-\frac{m}{2}) -
\frac{N(d-\frac{m}{2})}{2})} \Gamma_{(\tau)}^{(N)}(k_{i (\tau)}).
\end{equation}
We then define the dimension of the field $\phi$ as
\begin{equation}
\Gamma_{(1)}^{(N)}(\rho_{\tau}k_{i (\tau)}) = \rho_{\tau}^{\tau [(d-\frac{m}{2}) - N d_{\phi (\tau)}]}
\Gamma_{(1)}^{(N)}(k_{i}),
\end{equation}
such that in the free theory $d_{\phi}^{0} = \frac{d - \frac{m}{2}}{2} - 1$
is the naive dimension of the field. At the fixed point, the naive
dimension is modified due to the presence of interactions, such that
the nontrivial effect is the appearance of the anomalous dimension, i.e.,
$d_{\phi (\tau)}= \frac{d - \frac{m}{2}}{2} - 1 + \frac{\eta_{\tau}}{2 \tau}$. When $N=2$, this identifies the anomalous dimensions of the field in the
anisotropic situation, namely that associated to the change in the external
momenta scale perpendicular to the competing axes
$\eta_{1} \equiv \eta_{L2} = \gamma_{\phi (1)}(u_{1}^{*})$ as well as the
other corresponding to the change in the external
momenta parallel to the competition subspace
$\eta_{2} \equiv \eta_{L4} = \gamma_{\phi (2)}(u_{2}^{*})$.

This can be easily generalized to include $L$ insertions of  $\phi^{2}$
operators in quite an analogous way, such that the RG equations at
the fixed point lead to the solution ($(N,L) \neq (0,2)$) :

\begin{eqnarray}
&\Gamma_{R (\tau)}^{(N,L)} (\rho k_{i (\tau)}, \rho p_{i (\tau)}, u_{\tau}^{*},
\kappa_{\tau}) = \rho_{\tau}^{\tau [N + (d - \frac{m}{2}) - \frac{N(d - \frac{m}{2})}{2} -2L] -
\frac{N \gamma_{\phi (\tau)}^{*}}{2} + L \gamma_{\phi^{2} (\tau)}^{*}}
\nonumber\\
& \qquad \times \Gamma_{R (\tau)}^{(N,L)} (k_{i (\tau)}, p_{i (\tau)}, u_{\tau}^{*}, \kappa_{\tau}).
\end{eqnarray}
Thus, if we write at the fixed point
\begin{equation}
\Gamma_{R (\tau)}^{(N,L)} (\rho k_{i (\tau)}, \rho p_{i (\tau)}, u_{\tau}^{*},
\kappa_{\tau}) = \rho_{\tau}^{\tau [(d-\frac{m}{2}) - N d_{\phi}] + L d_{\phi^{2}}}
\Gamma_{R (\tau)}^{(N,L)} (k_{i (\tau)}, p_{i (\tau)}, u_{\tau}^{*}, \kappa_{\tau}),
\end{equation}
the anomalous dimensions of the insertions of $\phi^{2}$ operators are given
by $d_{\phi^{2}} = -2 \tau + \gamma_{\phi^{2} (\tau)}(u_{\tau}^{*})$ and
will be related to the critical exponents $\nu_{1} \equiv \nu_{L2}$ and
$\nu_{2} \equiv \nu_{L4}$ as we shall see in a moment.

In order to find the scaling relations we must go away from the Lifshitz
critical temperature ($t \neq 0$) staying, however, at the critical region
$\delta_{0} = 0$ \cite{Le2}. Above the Lifshitz critical
temperature, the renormalized vertex parts for
$t\neq 0$ can be expanded as a power series in $t$ around those renormalized
vertices at $t=0$, provided $N\neq 0$. We can now apply the differential
operators
\begin{equation}
O_{\tau} = \kappa_{\tau} \frac{\partial}{\partial \kappa_{\tau}} +
\beta_{\tau}\frac{\partial}{\partial u_{\tau}}
- \frac{1}{2} N \gamma_{\phi (\tau)}(u_{\tau}) +
\gamma_{\phi^{2} (\tau)}(u_{\tau}) t \frac{\partial}{\partial t}
\end{equation}
to $ \Gamma_{R (\tau)}^{(N)}(k_{i (\tau)})$ such that we find
\begin{eqnarray}
O_{\tau} \Gamma_{R (\tau)}^{(N)} (k_{i (\tau)}, t, u_{\tau}^{*}, \kappa_{\tau}) &=&
\sum_{L=0}^{\infty} \frac{t^{L}}{L!} [\kappa_{\tau} \frac{\partial}{\partial \kappa_{\tau}} +
\beta_{\tau}\frac{\partial}{\partial u_{\tau}}
- \frac{1}{2} N \gamma_{\phi (\tau)}(u_{\tau}) + L \gamma_{\phi^{2} (\tau)}]
\\
&& \qquad \times \;\; \Gamma_{R (\tau)}^{(N,L)} (k_{i (\tau)}, p_{i (\tau)}, u_{\tau}^{*}, \kappa_{\tau})\nonumber,
\end{eqnarray}
The result is that each term in the sum vanishes because of the RGE for
$\Gamma_{R (\tau)}^{(N,L)} (k_{i (\tau)}, p_{i (\tau)}, u_{\tau}^{*},
\kappa_{\tau})$. Hence, we obtain
\begin{equation}
[\kappa_{\tau} \frac{\partial}{\partial \kappa_{\tau}} +
\beta_{\tau}\frac{\partial}{\partial u_{\tau}}
- \frac{1}{2} N \gamma_{\phi (\tau)}(u_{\tau}) +
\gamma_{\phi^{2} (\tau)}(u_{\tau}) t \frac{\partial}{\partial t}]
\Gamma_{R (\tau)}^{(N)} (k_{i (\tau)}, t, u_{\tau}^{*}, \kappa_{\tau}) = 0.
\end{equation}
The solution is a homogeneous function of the product of $k_{i (\tau)}$
(to some power) and $t$ solely at the fixed point $u_{\tau}^{*}$. As the value
of $u_{\tau}$ is fixed at $u_{\tau}^{*}$, we shall omit it from the notation
of this section from now on. It is given by:
\begin{equation}
\Gamma_{R (\tau)}^{(N)} (k_{i (\tau)}, t, \kappa_{\tau})=
\kappa_{\tau}^{\frac{N \gamma_{\phi (\tau)}^{*}}{2}}
F_{(\tau)}^{(N)}(k_{i (\tau)},\kappa_{\tau} t^{\frac{-1}{\gamma_{\phi^{2} (\tau)}^{*}}}) .
\end{equation}
If we define $\theta_{\tau} = -\gamma_{\phi^{2} (\tau)}^{*}$, and using
dimensional analysis, we find
\begin{eqnarray}
\Gamma_{R (\tau)}^{(N)} (k_{i (\tau)}, t, \kappa_{\tau}) =&&
\rho_{\tau}^{\tau [N + (d- \frac{m}{2}) - \frac{N}{2}(d - \frac{m}{2})] -\frac{N}{2}
\eta_{\tau}}
\kappa_{\tau}^{\frac{N}{2} \eta_{\tau}} \nonumber\\
&&F_{(\tau)}^{(N)}(\rho_{\tau}^{-1} k_{i (1)},(\rho^{-1}\kappa_{\tau})
(\rho^{-2 \tau}t)^{\frac{-1}{\theta_{\tau}}} ) .
\end{eqnarray}

By choosing
$\rho_{\tau} = \kappa_{\tau} (\frac{t}{\kappa_{\tau}^{2 \tau}})^{\frac{1}{\theta_{\tau} + 2 \tau}}$, and
replacing back in (29), the vertex function depends only on the combination
$k_{i (\tau)} \xi_{\tau}$ apart from a power of $t$. Since the correlation
lengths $\xi_{\tau}$ are proportional to $t^{- \nu_{\tau}}$, we can identify
the critical exponents $\nu_{\tau}$ as
\begin{equation}
\nu_{\tau}^{-1} = 2 \tau + \theta_{\tau}^{*} = 2 \tau - \gamma_{\phi^{2} (\tau)}^{*} .
\end{equation}
According to our conventions, these equations are equivalent to the following
scaling relations
\begin{mathletters}
\begin{eqnarray}
\nu_{L2}^{-1} &=& 2 - \gamma_{\phi^{2} (1)}^{*},\\
\nu_{L4}^{-1} &=& 4 - \gamma_{\phi^{2} (2)}^{*}.
\end{eqnarray}
\end{mathletters}

As a matter of convenience, we could have defined alternatively the function
\begin{equation}
\bar{\gamma}_{\phi^{2} (\tau)}(u_{\tau}) = - \beta_{\tau}
\frac{\partial ln (Z_{\phi^{2} (\tau)}Z_{\phi (\tau)}) }{\partial u_{\tau}}.
\end{equation}
In that case we would have found the equivalent formulae
\begin{mathletters}
\begin{eqnarray}
\nu_{L2}^{-1} &=& 2 - \eta_{L2} - \bar{\gamma}_{\phi^{2} (1)}(u_{1}^{*}),\\
\nu_{L4}^{-1} &=& 4 - \eta_{L4} -\bar{\gamma}_{\phi^{2} (2)}^{*}.
\end{eqnarray}
\end{mathletters}
Hence, at the fixed point all correlation functions (not including composite
operators) scale at $T>T_{L}$,
since they are functions of  $k_{i (\tau)} \xi_{\tau}$ only.
For $N=2$ we choose
$\rho_{\tau} = k_{(\tau)}$, the external momenta. Then
$\Gamma_{R(\tau)}^{(2)}(k_{(\tau)}, t, \kappa_{\tau}) =
k^{2 \tau - \eta_{\tau}}
\kappa_{\tau}^{\eta_{\tau}} f(k_{(\tau)} \xi_{\tau})$.
The critical situation is characterized when $\xi_{\tau} \rightarrow \infty$
and $k_{(\tau)} \rightarrow 0$ such that
$f(k_{(\tau)} \xi_{\tau}) \rightarrow Constant$.
Defining $f_{\tau} =  (k_{(\tau)} \xi_{\tau})^{2 \tau - \eta_{\tau}}
f(k_{(\tau)} \xi_{\tau})$, we have
\begin{equation}
\Gamma_{R(\tau)}^{(2)}(k_{(\tau)}, t, \kappa_{\tau}) = (k_{(\tau)} \xi_{\tau})^{2 \tau - \eta_{\tau}}
\kappa_{\tau}^{\eta_{\tau}} f_{\tau}(k_{(\tau)} \xi_{\tau}).
\end{equation}
The susceptibility  is
proportional to $ t^{-\gamma_{\tau}}$ as $k_{(\tau)} \rightarrow 0$.
Thus, since $\Gamma_{R (\tau)}^{(2)} = \chi_{(\tau)}^{-1}$,
we can identify the susceptibility critical exponents
\begin{equation}
\gamma_{\tau} = \nu_{\tau} (2 \tau  - \eta_{\tau}).
\end{equation}
These relations are equivalent to the relations:
\begin{equation}
\gamma_{L2} = \nu_{L2} (2   - \eta_{L2}),
\end{equation}
\begin{equation}
\gamma_{L4} = \nu_{L4} (4   - \eta_{L4}).
\end{equation}

The specific heat exponents can be obtained by analysing the RG equation for
$\Gamma_{R (\tau)}^{(0,2)}$ above $T_{L}$ at the fixed point, i.e.
\begin{equation}
(\kappa_{\tau} \frac{\partial}{\partial \kappa_{\tau}} +
\gamma_{\phi^2 (\tau)}^{*} (2 + t\frac {\partial}{\partial t}))
\Gamma_{R (\tau)}^{(0,2)} =
(\kappa_{\tau}^{-2 \tau})^{\frac{\epsilon_{L}}{2}} B_{\tau}(u_{\tau}^{*}) ,
\end{equation}
where $B_{\tau}(u_{\tau}^{*})$ is given by
\begin{equation}
(\kappa_{\tau}^{-2 \tau})^{\frac{\epsilon_{L}}{2}}
B_{\tau}(u_{\tau}^{*}) = - Z_{\phi^{2}(\tau)}^{2}
\kappa_{\tau} \frac{\partial}{\partial \kappa_{\tau}}
\Gamma_{(\tau)}^{(0,2)}(Q_{(\tau)}; -Q_{(\tau)}, \lambda_{\tau})
|_{Q_{(\tau)}^{2}=\kappa_{\tau}^{2}}.
\end{equation}
It is a inhomogeneous part which has no dependence in the reduced temperature
$t$. The bare vertex function $\Gamma_{(\tau)}^{(0,2)}$ is
calculated as before in the limit $\Lambda_{\tau} \rightarrow \infty$, with
a fixed bare coupling constant, which renders $B_{\tau}(u_{\tau}^{*})$
finite in
this limit when $(d - \frac{m}{2}) = 4$. This renormalized vertex part
consists of the addition of the homogeneous (temperature dependent) and
inhomogeneous pieces. The general discussion given so far for the vertex part
$\Gamma_{R (\tau)}^{(N,L)}$ will be useful to determine the homogeneous part
of the solution. Indeed, at the fixed point the obvious generalization of the
solution for $\Gamma_{R(\tau)}^{(N,L)}$ is given by:
\begin{equation}
\Gamma_{R (\tau)}^{(N,L)} (p_{i (\tau)}, Q_{i (\tau)},  t, \kappa_{\tau}) =
\kappa_{\tau}^{\frac{1}{2} N \gamma_{\phi(\tau)}^{*} -
L \gamma_{\phi^{2}(\tau)}^{*}} F_{\tau}^{(N,L)}(p_{i (\tau)}, Q_{i (\tau)},
\kappa_{\tau} t^{\frac{-1}{\gamma_{\phi^{2} (\tau)}^{*}}}) .
\end{equation}

Therefore, the temperature dependent homogeneous part for
$\Gamma_{R(\tau),h}^{(0,2)}$ will scale at the fixed point as:
\begin{equation}
\Gamma_{R (\tau),h}^{(0,2)}(Q_{(\tau)}, -Q_{(\tau)}, t, \kappa_{\tau}) =
\kappa_{\tau}^{- 2 \gamma_{\phi^{2}(\tau)}^{*}} F_{\tau}^{(0,2)}
(Q_{(\tau)},- Q_{(\tau)},
\kappa_{\tau} t^{\frac{-1}{\gamma_{\phi^{2} (\tau)}^{*}}}) .
\end{equation}
This will be identified with the specific heat at zero external
momentum insertion $Q_{(\tau)}=0$. Using the results of our dimensional
analysis
\begin{equation}
\Gamma_{R (\tau),h}^{(0,2)})(Q_{(\tau)}, -Q_{(\tau)}, t, \kappa_{\tau}) =
\rho_{\tau}^{\tau[(d - \frac{m}{2}) - 4] + 2\gamma_{\phi^{2} (\tau)}^{*}}
\Gamma_{R (\tau),h}^{(0,2)}(\rho_{\tau}^{-1}Q_{(\tau)},
- \rho_{\tau}^{-1}Q_{(\tau)},
\rho_{\tau}^{-2 \tau} t,\rho_{\tau}^{-1} \kappa_{\tau}) ,
\end{equation}
and replacing this into the solution at the fixed point, we find
\begin{eqnarray}
&&\Gamma_{R (\tau),h}^{(0,2)}(Q_{(\tau)}, -Q_{(\tau)}, t, \kappa_{\tau}) =
\rho_{\tau}^{\tau [(d - \frac{m}{2}) - 4] + 2\gamma_{\phi^{2} (\tau)}^{*}}
\kappa_{\tau}^{- 2 \gamma_{\phi^{2}(\tau)}^{*}}\\
&& \qquad \times \;\; F_{\tau}^{(0,2)}(\rho_{\tau}^{-1} Q_{(\tau)},
-\rho_{\tau}^{-1} Q_{(\tau)},
\rho_{\tau}^{-1} \kappa_{\tau}(\rho_{\tau}^{-2 \tau} t)^{\frac{-1}{\gamma_{\phi^{2} (\tau)}^{*}}}) .\nonumber
\end{eqnarray}
Again we choose
$\rho_{\tau} = \kappa_{\tau} (\frac{t}{\kappa_{\tau}^{2 \tau}})^{\frac{1}{\theta_{\tau} + 2 \tau}}$. Substituting this in last equation, taking the limit
$Q_{(\tau)} \rightarrow 0$ and identifying the power of $t$ with the specific
heat exponent $\alpha_{\tau}$, we find:
\begin{equation}
\alpha_{\tau} = 2 - \tau (d-\frac{m}{2})\nu_{\tau} .
\end{equation}
Let us analyse the inhomogeneous part. First, take $Q_{(\tau)}=0$. Second,
choose a particular solution of the form:
\begin{equation}
C_{p}(u_{\tau}) = (\kappa_{\tau}^{2 \tau})^{\frac{- \epsilon_{L}}{2}}
\tilde{C}_{p}(u_{\tau}).
\end{equation}
Replace this into the RG equation for $\Gamma_{R (\tau)}^{(0,2)}$ at the
fixed point. Then, it is easy to obtain
\begin{equation}
C_{p}(u_{\tau}^{*}) = (\kappa_{\tau}^{2 \tau})^{\frac{- \epsilon_{L}}{2}}
\frac{\nu_{\tau}}{\nu_{\tau} \tau(d - \frac{m}{2}) -2} B_{\tau}(u_{\tau}^{*}).
\end{equation}
Collecting both terms we have the following general solution at the fixed
point:
\begin{equation}
\Gamma_{R (\tau)}^{(0,2)} = (\kappa_{\tau}^{-2 \tau})^{\frac{\epsilon_{L}}{2}}
(C_{\tau} (\frac{t}{\kappa_{\tau}^{2 \tau}})^{- \alpha_{\tau}} +
\frac{\nu_{\tau}}{\nu_{\tau} \tau(d - \frac{m}{2}) -2} B_{\tau}(u_{\tau}^{*})).
\end{equation}

Let us describe the situation for $T<T_{L}$. It can be illustrated for the
case of magnetic systems. The renormalized equation of
state relates the renormalized (1PI one-point vertex part) magnetic field
with the renormalized vertex parts for $t<0$ via a power series in the
magnetization $M$. One has:
\begin{equation}
H_{(\tau)}(t, M, u_{\tau}, \kappa_{\tau}) = \sum_{N=1}^{\infty} \frac{1}{N!} M^{N}
\Gamma_{R (\tau)}^{1+N}(k_{i (\tau)} = 0; t, u_{\tau}, \kappa_{\tau}),
\end{equation}
where the zero momentum limit must be taken after realizing the summation.
The magnetic field satisfies the following RG equation:
\begin{equation}
(\kappa_{\tau} \frac{\partial}{\partial \kappa_{\tau}} +
\beta_{\tau}\frac{\partial}{\partial u_{\tau}}
- \frac{1}{2} N \gamma_{\phi (\tau)}(N + M \frac{\partial}{\partial M}) +
\gamma_{\phi^{2} (\tau)} t \frac{\partial}{\partial t})
H_{(\tau)}(t, M, u_{\tau}, \kappa_{\tau}) = 0 .
\end{equation}
At the fixed point we have the following form for the equation of state:
\begin{equation}
H_{(\tau)}(t, M, \kappa_{1}) = \kappa_{\tau}^{\frac{\eta_{\tau}}{2}}
h_{1 \tau}(\kappa_{\tau} M^{\frac{2}{\eta_{\tau}}}, \kappa_{\tau} t^{\frac{-1}
{\gamma_{\phi^{2} (\tau)}}}).
\end{equation}
Once again, we use dimensional analysis arguments to obtain the following
expression under a flow in the momenta :
\begin{equation}
H_{(\tau)}(t, M, \kappa_{\tau}) = \rho_{\tau}^{\tau [\frac{d - \frac{m}{2}}{2} + 1]} H_{(\tau)}(\frac{t}{\rho_{\tau}^{2 \tau}}, \frac{M}{\rho_{\tau}^{2 \tau[\frac{d - \frac{m}{2}}{2} - 1]}},
\frac{\kappa_{\tau}}{\rho_{\tau}}) .
\end{equation}
We choose $\rho_{\tau}$ to be a power of $M$ such that:
\begin{equation}
\rho_{\tau} = \kappa_{\tau} [\frac{M}{\kappa_{\tau}^{\frac{\tau}{2}[(d - \frac{m}{2}) - 2]}}]^{\frac{2}
{\tau [(d - \frac{m}{2}) - 2] + \eta_{\tau}}} ,
\end{equation}
and from the scaling form of the equation of state
$H_{(\tau)}(t, M) =
M^{\delta_{\tau}} f(\frac{t}{M^{\frac{1}{\beta_{\tau}}}})$, we
obtain the remaining scaling laws after the identifications
$\delta_{1} = \delta_{L2}, \beta_{1} = \beta_{L2}$, $\delta_{2} = \delta_{L4}, \beta_{2} = \beta_{L4}$,  :
\begin{mathletters}
\begin{eqnarray}
\delta_{L2} &=& \frac{(d-\frac{m}{2}) + 2 - \eta_{L2}}{(d-\frac{m}{2}) - 2 + \eta_{L2}},\\
\beta_{L2} &=& \frac{1}{2} \nu_{L2} ((d-\frac{m}{2}) - 2 + \eta_{L2}),\\
\delta_{L4} &=& \frac{2(d-\frac{m}{2}) + 4 - \eta_{L4}}{2(d-\frac{m}{2}) - 4 + \eta_{L4}},\\
\beta_{L4} &=& \frac{1}{2} \nu_{L4} (2(d-\frac{m}{2}) - 4 + \eta_{L4}),\\
\end{eqnarray}
\end{mathletters}
which imply the Widom $\gamma_{L2} = \beta_{L2} (\delta_{L2} -1)$ and
Rushbrook $\alpha_{L2} + 2 \beta_{L2} + \gamma_{L2} = 2$ relations for
directions perpendicular to the competition axes. These relations are also
valid for directions along the competing axes. So far, the effect of
considering this new dimensional role played by the momenta
scale along the competing quartic subspace, together with the definitions of
the critical theories either at vanishing quartic or quadratic external
momenta have induced two independent set of scaling relations for the
critical exponents. Nevertheless, when performing the diagramatic
perturbative expansion, we shall find out that some of these exponents
are not independent.

There is one curious fact relating these scaling relations and those
first obtained by Hornreich et al. \cite{Ho-Lu-Sh}.
If $\nu_{L4} = \frac{1}{2} \nu_{L2}$
and $\eta_{L4} = 2 \eta_{L2}$ to all orders in perturbation theory, the
hyperscaling (Josephson) relations are the same in either formulation. In
the formulation presented here this feature will be found in Sec. VI from
the perturbative analysis up to the loop order considered in this paper.
In addition, one obtains
\begin{mathletters}
\begin{eqnarray}
&\gamma_{L4} = \gamma_{L2} = \gamma_{L}, \\
&\beta_{L4} = \beta_{L2} = \beta_{L}, \\
&\delta_{L4} = \delta_{L2} = \delta_{L}, \\
&\alpha_{L4} = \alpha_{L2} = \alpha_{L}.
\end{eqnarray}
\end{mathletters}
In that case, there is a complete equivalence among the scaling
relations in both formulations. It is worth to emphasize that the advantage
of the approach presented in this section is the splitting of the
scaling laws into independent renormalization group flows parallel and
perpendicular to the competition axes. Then, instead of claiming that
the two momenta scales corresponding to the components perpendicular
and parallel to the competing axes are equal \cite{Ho-Lu-Sh}, the most
important conclusion in our approach with two coupling constants is that they
will flow to the same fixed point in the critical regime as will be
shown later.

We can make a comparison among our scaling relations below
$T_{L}$ with the ones obtained by Mergulh\~ao and
Carneiro. In their work, they defined the space
dimension of the Lifshitz system $D\equiv \frac{d_{\alpha}}{2} + d_{\beta}$,
where $d_{\alpha} = m$ and $d_{\beta} = d-m$ (see Eq.(25)
in \cite{Mergulho1}). Therefore, $D=(d-\frac{m}{2})$, the same effective
space dimension as ours.

The trouble is in the introduction of $\sigma$.
The normalization conditions defined through Eqs.(20)-(24) of
Ref. \cite {Mergulho1} mix $\sigma$ with the two external momenta scales
in a nontrivial way. They recognized, however, that the normalization
conditions they defined are {\it independent} of $\sigma$. Thus if
one makes the choice $\gamma_{\sigma}^{*}=0$, Eq. (58) in
Ref. \cite {Mergulho1} is just the same as the one obtained here for
$\delta_{L}$, and Eq.(60) in Ref. \cite {Mergulho1} is equal to that
obtained here for $\beta_{L}$. The Eqs. (46)-(51) in their article taken
together with $\gamma_{\sigma}^{*}=0$ yields trivially
$\nu_{L4} = \frac{1}{2} \nu_{L2}$ and $\eta_{L4} = 2 \eta_{L2}$.
Last but not least, if one takes the
bare $\sigma_{0}$ dimensionless, as was done by those authors, the whole
argument is invalidated since its {\it dimensionfulness} was assured from the
beginning of the discussion in the regulation of the free critical
propagator. Introducing $\sigma$ is consistent provided it is considered as
a dimensionful ammount in all stages of the calculation. In other words,
$\sigma$ is not required at all, since the flow in $\sigma$ can be absorbed
in the quartic momenta scale using our dimensional redefinition.

\section{Renormalization group for the isotropic behavior}
The procedure to analyse the isotropic case is quite analogous. Some care
must be taken. Whenever $\tau$ appears as a subscript, like in a quantity
$A_{\tau}$, one sets $\tau =3$ in order to be consistent with the notation
employed so far. The dimensional analysis is a bit
different. The volume element in momentum space is again
$d^{d-m}q d^{m}k$. Whenever $d=m$, the volume element is
now $d^{m}k$. As before, $[k] = M^{\frac{1}{2}}$. Accordingly, the
volume element has dimension $[d^{m}k] = M^{\frac{m}{2}}$. The
dimension of the field in mass units is
$[\phi] = M^{\frac{m}{4} - 1}$. When the conserving $\delta$-function
is removed the 1PI vertex parts have dimensions
$[\Gamma^{(N)}(k_{i})] = M^{N + \frac{m}{2} - N \frac{m}{4}}$.
Then, make the continuation $m=8-\epsilon_{L}$. The coupling
constant has dimension
$\lambda_{3} = M^{\frac{8-m}{2}}= M^{\frac{\epsilon_{L}}{2}}$.
In terms of dimensionless
quantities, one has the renormalized
$g_{3} = u_{3} (\kappa_{3}^{4})^{\frac{\epsilon_{L}}{4}}$ and
bare $\lambda_{3} = u_{03} (\kappa_{3}^{4})^{\frac{\epsilon_{L}}{4}}$
coupling constants, respectively. Again, the functions
\begin{mathletters}
\begin{eqnarray}
\beta_{3} &=& (\kappa_{3}\frac{\partial u_{3}}{\partial \kappa_{3}})\\
\gamma_{\phi (3)}(u_{3})  &=& \beta_{3}
\frac{\partial ln Z_{\phi (3)}}{\partial u_{3}}\\
\gamma_{\phi^{2} (3)}(u_{3}) &=& - \beta_{3}
\frac{\partial ln Z_{\phi^{2} (3)}}{\partial u_{3}}
\end{eqnarray}
\end{mathletters}
are calculated, as before, at fixed bare coupling $\lambda_{3}$.
The beta function can be expressed in terms of dimensionless quantities as
$\beta_{3} = -  \epsilon_{L}(\frac{\partial ln u_{0 3}}{\partial u_{3}})^{-1}$. One should notice that the beta function for the isotropic case does not
possess the overall factor of 2 present in the anisotropic beta function
$\beta_{2}$ obtained from renormalization group transformations over the
quartic momenta scale $\kappa_{2}$. This feature is a strong nonperturbative
suggestion that the isotropic critical observables cannot be obtained from
the anisotropic ones and vice-versa.

Under a flow in the quartic momenta, from our dimensional analysis,
the dimensional redefinition of the momenta along the competing axes
results in  an effective space dimension for the isotropic case, i.e.,
$(\frac{m}{2})$. Thus, we find
the following behavior for the 1PI vertex parts $\Gamma_{R (3)}^{(N)}$ under
a flow in the external momenta:
\begin{eqnarray}
&\Gamma_{R (3)}^{(N)} (\rho_{3} k_{i}, u_{3}, \kappa_{3}) =
\rho_{3}^{2[N + \frac{m}{2} - N \frac{m}{4}]}
exp[-\frac{N}{2} \int_{1}^{\rho_{3}} \gamma_{\phi (3)}(u_{3}(\rho_{3}))
\frac{{d x_{3}}}{x_{3}}] \\
& \times \;\Gamma_{R (3)}^{(N)} (k_{i}, u_{3}(\rho_{3}),
\kappa_{3} \rho_{3})\nonumber.
\end{eqnarray}
Notice that we put aside the notation $k_{i (\tau)}$, etc., used in the
anisotropic analysis in favor of $k_{i}$, etc. since there is only one
quartic momenta scale in the isotropic case. At the fixed point, we also
have a simple scaling property for the vertex
parts $\Gamma_{R (3)}^{(N)}$, namely:

\begin{eqnarray}
&\Gamma_{R (3)}^{(N)} (\rho_{3} k_{i}, u_{3}^{*}, \kappa_{3}) =
\rho_{3}^{2[N + \frac{m}{2} - N \frac{m}{4}] -
\frac{N}{2} \gamma_{\phi (3)}(u_{3}^{*})}\\
& \times \Gamma_{R (3)}^{(N)} (k_{i}, u_{3}^{*},\kappa_{3})\nonumber.
\end{eqnarray}

For $N=2$, we have
\begin{equation}
\Gamma_{R (3)}^{(2)} (\rho_{3} k, u_{3}^{*}, \kappa_{3}) =
\rho_{3}^{4 - \gamma_{\phi (3)}(u_{3}^{*})}\Gamma_{R (3)}^{(2)} (k, u_{3}^{*}, \kappa_{3} ).
\end{equation}
We can now identify
$\eta_{L4} \equiv \eta_{3} = \gamma_{\phi (3)}(u_{3}^{*})$ as the
anomalous dimension for the isotropic case. This is the analogue of
the analysis we performed for the anisotropic case. In the free theory
$d_{\phi}^{0} = \frac{\frac{m}{2}}{2} - 1$
is the naive dimension of the field. At the isotropic fixed point, the naive
dimension is modified due to the presence of interactions, such that
$d_{\phi}= \frac{\frac{m}{2}}{2} - 1 + \frac{\eta_{L4}}{4}$.
The generalization to include $L$ insertions of  $\phi^{2}$
operators is quite straightforward and can be written at the fixed point as
($(N,L) \neq (0,2)$) :

\begin{equation}
\Gamma_{R (3)}^{(N,L)} (\rho_{3} k_{i}, \rho_{3} p_{i}, u_{3}^{*},
\kappa_{3}) = \rho_{3}^{2 [N + \frac{m}{2} - \frac{N(\frac{m}{2})}{2} -2L] -
\frac{N \gamma_{\phi (3)}^{*}}{2} + L \gamma_{\phi^{2} (3)}^{*}}
\Gamma_{R (3)}^{(N,L)} (k_{i}, p_{i}, u_{3}^{*}, \kappa_{3}).
\end{equation}
Thus, if we write at the fixed point
\begin{equation}
\Gamma_{R (3)}^{(N,L)} (\rho_{3} k_{i}, \rho_{3} p_{i}, u_{3}^{*},
\kappa_{3}) = \rho_{3}^{ \frac{m}{2} - N d_{\phi} + L d_{\phi^{2}}}
\Gamma_{R (3)}^{(N,L)} (k_{i}, p_{i}, u_{3}^{*}, \kappa_{3}),
\end{equation}
the anomalous dimension of the insertions of $\phi^{2}$ operators is given by
$d_{\phi^{2}} = -4 + \gamma_{\phi^{2} (3)}(u_{3}^{*})$.

Above the Lifshitz critical temperature, the renormalized vertex parts for
$t\neq 0$ can be expanded as a power series in $t$ around those renormalized
vertices at $t=0$, provided $N\neq 0$. We can now apply the differential
operators
\begin{equation}
O_{3} = \kappa_{3} \frac{\partial}{\partial \kappa_{3}} +
\beta_{3}\frac{\partial}{\partial u_{3}}
- \frac{1}{2} N \gamma_{\phi (3)}(u_{3}) +
\gamma_{\phi^{2} (3)}(u_{3}) t \frac{\partial}{\partial t}
\end{equation}
to $ \Gamma_{R (3)}^{(N)}(k_{i},t, u_{3}^{*}, \kappa_{3})$. The mechanism is
similar to that discussed in the anisotropic cases. This vertex part is a
power series on $t$, with each individual coefficient vanishing by making use
of the RGE for
$\Gamma_{R (3)}^{(N,L)} (k_{i}, p_{i}, u_{3}^{*}, \kappa_{3})$. Then,
we find
\begin{equation}
[\kappa_{3} \frac{\partial}{\partial \kappa_{3}} +
\beta_{3}\frac{\partial}{\partial u_{3}}
- \frac{1}{2} N \gamma_{\phi (3)}(u_{3}) +
\gamma_{\phi^{2} (3)}(u_{3}) t \frac{\partial}{\partial t}]
\Gamma_{R (3)}^{(N)} (k_{i}, t, u_{3}^{*}, \kappa_{3}) = 0.
\end{equation}
The solution is a homogeneous function of the product of $k_{i}$
(to some power) and $t$ only at the fixed point $u_{3}^{*}$.
The solution reads:
\begin{equation}
\Gamma_{R (3)}^{(N)} (k_{i (3)}, t, u_{3}^{*}, \kappa_{3})=
\kappa_{3}^{\frac{N \gamma_{\phi (3)}^{*}}{2}}
F_{(3)}^{(N)}(k_{i},\kappa_{3} t^{\frac{-1}{\gamma_{\phi^{2} (3)}^{*}}}) .
\end{equation}
All the exponents generated by the renormalization group flow along the
scale $\kappa_{3}$ will be denoted by a corresponding $L4$ subscript. If we
define $\theta_{3} = -\gamma_{\phi^{2} (3)}^{*}$, one can use dimensional
analysis to obtain
\begin{eqnarray}
&\Gamma_{R (3)}^{(N)} (k_{i}, t, \kappa_{3}) =
\rho_{3}^{2 [N + \frac{m}{2} - \frac{N}{2} \frac{m}{2}] -\frac{N}{2}
\eta_{L4}}
\kappa_{3}^{\frac{N}{2} \eta_{L4}} \nonumber\\
& \times \; F_{(3)}^{(N)}(\rho_{3}^{-1} k_{i},(\rho_{3}^{-1}\kappa_{3})
(\rho_{3}^{-4}t)^{\frac{1}{\theta_{3}}} ) .
\end{eqnarray}

One can choose
$\rho_{3} = \kappa_{3} (\frac{t}{\kappa_{3}^{4}})^{\frac{1}{\theta_{3} + 4}}$, and replacing it in (64), the vertex function depends only on the combination
$k_{i} \xi_{L4}$ apart from a power of $t$. As $\xi_{L4}$ is proportional to
$t^{-\nu_{L4}}$ the critical exponent
$\nu_{3} = \nu_{L4}$ can be identified as
\begin{equation}
\nu_{L4}^{-1} = 4  + \theta_{3}^{*} = 4 - \gamma_{\phi^{2} (3)}^{*} .
\end{equation}

Again it is convenient to define the function
\begin{equation}
\bar{\gamma}_{\phi^{2} (3)}(u_{3}) = - \beta_{3}
\frac{\partial ln (Z_{\phi^{2} (3)}Z_{\phi (3)}) }{\partial u_{3}}.
\end{equation}
Then, one can easily find the following relation
\begin{equation}
\nu_{L4}^{-1} = 4 - \eta_{L4} - \bar{\gamma}_{\phi^{2} (3)}(u_{3}^{*}).
\end{equation}
At the fixed point all correlation functions (not including composite
operators) scale at $T>T_{L}$, since they are functions of
$k_{i} \xi_{L4}$ only. For $N=2$ we choose
$\rho_{3} = k$, the external momenta. The two-point vertex part can be
written in the form $\Gamma_{R(3)}^{(2)}(k, t, \kappa_{3}) = k^{4 - \eta_{L4}}
\kappa_{3}^{\eta_{L4}} f(k \xi_{L4})$. The main point is that when
 $\xi_{L4} \rightarrow \infty$ and $k \rightarrow 0$, simultaneously, then
$f(k \xi_{L4}) \rightarrow Constant$.
Defining $f_{3} =  (k \xi_{L4})^{4 - \eta_{L4}} f(k \xi_{L4})$, we have
$\Gamma_{R(3)}^{(2)}(k, t, \kappa_{3}) = (k \xi_{L4})^{4 - \eta_{L4}}
\kappa_{3}^{\eta_{L4}} f_{3}(k \xi_{L4})$. The susceptibility  is
proportional to $ t^{-\gamma_{L4}}$ as $k_{i} \rightarrow 0$. Since
$\Gamma_{R}^{(2)} = \chi^{-1}$, the susceptibility critical
exponent is given by
\begin{equation}
\gamma_{L4} = \nu_{L4} (4  - \eta_{L4}).
\end{equation}

The scaling relation for the specific heat exponent can be found from
the RG equation for
$\Gamma_{R (3)}^{(0,2)}$ above $T_{L}$ at the fixed point, namely
\begin{equation}
(\kappa_{3} \frac{\partial}{\partial \kappa_{3}} +
\gamma_{\phi^2 (3)}^{*} (2 + t\frac {\partial}{\partial t}))
\Gamma_{R (3)}^{(0,2)} =  (\kappa_{3}^{-2 })^{\frac{\epsilon_{L}}{2}}
B_{3}(u_{3}^{*}) ,
\end{equation}
where
\begin{equation}
(\kappa_{3}^{-4})^{\frac{\epsilon_{L}}{4}}
B_{3}(u_{3}^{*}) = - Z_{\phi^{2}(3)}^{2}
\kappa_{3} \frac{\partial}{\partial \kappa_{3}}
\Gamma_{(3)}^{(0,2)}(Q; -Q, \lambda_{3}),
\end{equation}
is the inhomogeneous part which does not depend on $t$. Recall that the bare
vertex function $\Gamma_{(3)}^{(0,2)}$ is
calculated as before in the limit $\Lambda_{3} \rightarrow \infty$, with
a fixed bare coupling constant, which renders $B_{3}(u_{3}^{*})$ finite in
this limit when $m= 8$. This renormalized vertex part is made out of the
addition of the homogeneous (temperature dependent) and
inhomogeneous pieces. The general discussion given so far for the vertex part
$\Gamma_{R (3)}^{(N,L)}$ is helpful to obtain the homogeneous part of
the solution. At the fixed point we have the following generalization of the
solution for $\Gamma_{R(3)}^{(N,L)}$:
\begin{equation}
\Gamma_{R (3)}^{(N,L)}(p_{i}, Q_{i},  t, \kappa_{3}) =
\kappa_{3}^{\frac{1}{2} N \gamma_{\phi(3)}^{*} -
L \gamma_{\phi^{2}(3)}^{*}} F_{3}^{(N,L)}(p_{i}, Q_{i},
\kappa_{3} t^{\frac{-1}{\gamma_{\phi^{2} (3)}^{*}}}) .
\end{equation}
The temperature dependent homogeneous part for
$\Gamma_{R(3),h}^{(0,2)}$ scales at the fixed point, i.e.,
\begin{equation}
\Gamma_{R (3),h}^{(0,2)}(Q, -Q, t, \kappa_{3}) =
\kappa_{3}^{- 2 \gamma_{\phi^{2}(3)}^{*}} F_{3}^{(0,2)}(Q,- Q,
\kappa_{3} t^{\frac{-1}{\gamma_{\phi^{2} (3)}^{*}}}) .
\end{equation}
This vertex function is to be identified with the specific heat at
zero external momentum insertion $Q=0$. Using our dimensional analysis
one finds:
\begin{equation}
\Gamma_{R (3),h}^{(0,2)}(Q, -Q, t, \kappa_{3}) =
\rho^{2[\frac{m}{2} - 4] + 2\gamma_{\phi^{2} (3)}^{*}}
\Gamma_{R (3),h}^{(0,2)}(\rho_{3}^{-1}Q, - \rho_{3}^{-1}Q,
\rho_{3}^{-4} t,\rho_{3}^{-1} \kappa_{3}) .
\end{equation}
Substitution of this equation into the solution at the fixed point yields
\begin{equation}
\Gamma_{R (3),h}^{(0,2)}(Q, -Q, t, \kappa_{3}) =
\rho_{3}^{2[\frac{m}{2} - 4] + 2\gamma_{\phi^{2} (3)}^{*}}
\kappa_{3}^{- 2 \gamma_{\phi^{2}(3)}^{*}} F_{3}^{(0,2)}(\rho_{3}^{-1} Q,
-\rho_{3}^{-1} Q,
\rho_{3}^{-1} \kappa_{3}(\rho_{3}^{-4} t)^{\frac{-1}{\gamma_{\phi^{2} (3)}^{*}}}) .
\end{equation}
We make the choice
$\rho_{3} = \kappa_{3} (\frac{t}{\kappa_{3}^{4}})^{\frac{1}{\theta_{3} + 4}}$.
Replacing this into the last equation, taking the limit
$Q \rightarrow 0$ and identifying the power of $t$ with the specific heat
exponent $\alpha_{L4}$, we find:
\begin{equation}
\alpha_{L4} = 2 - m \nu_{L4} .
\end{equation}
The description of the inhomogeneous part is as follows. First, take $Q=0$.
Then, choose a particular solution of the form:
\begin{equation}
C_{p}(u_{3}) = (\kappa_{3}^{4})^{\frac{- \epsilon_{L}}{4}}
\tilde{C}_{p}(u_{3}).
\end{equation}
Now replace this into the RG equation for $\Gamma_{R (3)}^{(0,2)}$ at the
fixed point. Therefore, one gets to
\begin{equation}
C_{p}(u_{3}^{*}) = (\kappa_{3}^{4})^{\frac{- \epsilon_{L}}{4}}
\frac{\nu_{L4}}{\nu_{L4} m -2} B_{3}(u_{3}^{*})).
\end{equation}
The general solution at the fixed point is just the sum of the two pieces,
and is given by
\begin{equation}
\Gamma_{R (3)}^{(0,2)} = (\kappa_{3}^{-4})^{\frac{\epsilon_{L}}{4}}
(C_{3} (\frac{t}{\kappa_{3}^{4}})^{- \alpha_{L4}} +
\frac{\nu_{L4}}{\nu_{L4} m -2} B_{3}(u_{3}^{*})).
\end{equation}

We now turn our attention to analyse the scaling relations when the system
is below the Lifshitz critical temperature $T<T_{L}$.
The renormalized magnetic field is related to the renormalized vertex
parts for $t<0$ and the magnetization $M$ through
\begin{equation}
H_{(3)}(t, M, u_{3}, \kappa_{3}) = \sum_{N=1}^{\infty} \frac{1}{N!} M^{N}
\Gamma_{R(3)}^{1+N}(k_{i} = 0; t, u_{3}, \kappa_{3}),
\end{equation}
where the zero momentum limit must be taken after realizing the summation.
The magnetic field satisfies the RG equation:
\begin{equation}
(\kappa_{3} \frac{\partial}{\partial \kappa_{3}} +
\beta_{3}\frac{\partial}{\partial u_{3}}
- \frac{1}{2} N \gamma_{\phi (3)}(u_{3})(N + M \frac{\partial}{\partial M}) +
\gamma_{\phi^{2} (3)} t \frac{\partial}{\partial t})
H_{(3)}(t, M, u_{1}, \kappa_{1}) = 0 .
\end{equation}
The equation of state at the fixed point reads:
\begin{equation}
H_{(3)}(t, M, \kappa_{3}) = \kappa_{3}^{\frac{\eta_{L4}}{2}}
h_{3}(\kappa_{3} M^{\frac{2}{\eta_{L4}}}, \kappa_{1} t^{\frac{-1}
{\gamma_{\phi^{2} (3)}}}).
\end{equation}
Dimensional analysis arguments lead to the following
expression under a flow in the external momenta:
\begin{equation}
H_{(3)}(t, M, \kappa_{3}) = \rho_{3}^{2 [\frac{m}{4} + 1]}\nonumber\\
\;\; H_{3}(\frac{t}{\rho_{3}^{4}}, \frac{M}{\rho_{3}^{2[\frac{m}{4} - 1]}},
\frac{\kappa_{3}}{\rho_{3}}) .
\end{equation}
The flow parameter $\rho_{3}$ is chosen to be a power of $M$ such that:
\begin{equation}
\rho_{3} = \kappa_{3} [\frac{M}{\kappa_{3}^{[\frac{m}{2} - 2]}}]^{\frac{2}
{m - 4 + \eta_{L4}}} ,
\end{equation}
and from the scaling form of the equation of state
$H_{(3)}(t, M) = M^{\delta_{L4}} f(\frac{t}{M^{\frac{1}{\beta_{L4}}}})$, we
obtain the following scaling laws:
\begin{mathletters}
\begin{eqnarray}
\delta_{L4} &=& \frac{m + 4 - \eta_{L4}}{m - 4 + \eta_{L4}}, \\
\beta_{L4} &=& \frac{1}{2} \nu_{L4} (m - 4 + \eta_{L4}),
\end{eqnarray}
\end{mathletters}
which imply the Widom $\gamma_{L4} = \beta_{L4} (\delta_{L4} -1)$ and
Rushbrook $\alpha_{L4} + 2 \beta_{L4} + \gamma_{L4} = 2$ relations.

The scaling relations for the anisotropic case Eqs.(11a),(12)
in Ref. \cite{Ho-Lu-Sh} for $d=m$ are consistent
with the isotropic case. Note, however, that this cannot be given a
rigorous meaning, for the appearance of $\nu_{L2}$ and $\eta_{L2}$ in the
equality in Eq.(11b) of Ref. \cite{Ho-Lu-Sh} invalidates the argument for
the isotropic case as these exponents are no longer meaningful. Notice that
the impossibility of finding scaling relations for the isotropic case in the
original framework \cite{Ho-Lu-Sh} is due to the lack of the
independent flow in the external momenta scale $\kappa_{3}$ along the
quartic subspace. In the early treatment \cite{Ho-Lu-Sh}, the quartic
momenta was not independent to be varied freely, but was fixed from
the variation of the quadratic scale. Without its free variation,
which is possible since this quartic term in the propagator does not
have the same canonical dimension as the quadratic one, no
renormalization group flows along the competing directions can
be defined whatsoever. Thus, this new renormalization group method permits
to go further in determining the Lifshitz critical universal properties of
the system for arbitrary $m$.

\section{The evaluation of Feynman integrals}

In order to calculate universal quantities like critical exponents, we
must calculate some Feynman integrals. The perturbative loop expansion
shall be our starting point with the $\epsilon_{L}= 4 + \frac{m}{2} - d$
being the perturbation parameter for the anisotropic situation. For the
isotropic case, the perturbation parameter is $\epsilon_{L} = 8 - m$.

We have to express the solution of the
Feynman diagrams in terms of $\epsilon_{L}$, resulting in the
$\epsilon_{L}$-expansion for the universal critical ammounts. Again,
there is also a very important difference among the anisotropic and
isotropic behaviors. From a technical viewpoint, the anisotropic behaviors
present two types of integration along the two momenta subspaces, whereas
in the isotropic situation there is only one subspace to be integrated
over. We shall treat them separately.

The anisotropic behavior is described using two different
approximations.  We shall briefly discuss the first analytical approximation
developed for evaluating higher-order Feynman diagrams which are needed
in the calculation of the critical exponents perpendicular to the competing
axes for the anisotropic Lifshitz behavior. It points out the necessity of
some sort of condition among the quartic loop momenta in different
subdiagrams, leading to the homogeneity of the integrals in the quadratic
external momenta scales. We employ the set of normalization conditions
with vanishing quartic external momenta as described in Sec.II.  This
piece of work was done in collaboration with L.C. de Albuquerque and the
details can be found in \cite{AL1,AL2}.

Nevertheless, with the renormalization group description presented here,
this approximation is not sufficient to describe the critical exponents
along the competing axes. It does not yield the solution of the integrals
as a homogeneous function of {\it both} quadratic and quartic external
momenta scales yet. The former approximation described above is then
generalized to obtain the solution of the integrals for {\it arbitrary}
quadratic and quartic external momenta scales. Using the new interpretation
for the momenta scale along the quartic direction given in the last two
sections, the calculation of these integrals is not a complicated task,
provided a certain condition among the quartic momenta is fulfilled.
With this new technique all the critical exponents in the anisotropic
cases are obtained. This picture can be considered the main result of
the present work.

The isotropic behavior can be developed along the same lines of the
latter approach to the anisotropic case. The condition among the quartic
momenta is also required in order to guarantee homogeneity of the Feynman
integrals in the quartic external momenta scale. The new approximation
is sufficient to complete the unified analytical description of the
Lifshitz critical behavior in its full generality, at least at the loop
order considered here as will be shown in this section.

In order to verify the renormalization scheme independence of the critical
exponents, it would be interesting to obtain the critical exponents using
more than one renormalization procedure. In fact, as will be proven later,
the use of normalization conditions or minimal subtraction of dimensional
poles yield the same critical exponents. Thus, we shall present the results
in the most appropriate form for calculating the critical exponents in these
two renormalization prescriptions.

\subsection{Anisotropic}

In order to calculate universal quantities like critical exponents, we
must calculate some Feynman integrals. We start by listing all the relevant
integrals which are necessary to find out the critical exponents.
These integrals are

\begin{equation}
I_2 =  \int \frac{d^{d-m}q d^{m}k}{[\bigl((k + K^{'})^{2}\bigr)^2 +
(q + P)^{2}] \left( (k^{2})^2 + q^{2}  \right)}\;\;\;,
\end{equation}
where $I_{2}$ is the one-loop integral contributing to the four-point
function,

\begin{equation}
I_{3} = \int \frac{d^{d-m}{q_{1}}d^{d-m}q_{2}d^{m}k_{1}d^{m}k_{2}}
{\left( q_{1}^{2} + (k_{1}^{2})^2 \right)
\left( q_{2}^{2} + (k_{2}^{2})^2 \right)
[(q_{1} + q_{2} + P)^{2} + \bigl((k_{1} + k_{2} + K')^{2}\bigr)^2]}\;\;,
\end{equation}
is the two-loop \lq\lq sunset'' Feynman diagram of the two-point
function,

\begin{eqnarray}
I_{4}\;\; =&& \int \frac{d^{d-m}{q_{1}}d^{d-m}q_{2}d^{m}k_{1}d^{m}k_{2}}
{\left( q_{1}^{2} + (k_{1}^{2})^2 \right)
\left( (P - q_{1})^{2} + \bigl((K' - k_{1})^{2}\bigr)^2  \right)
\left( q_{2}^{2} + (k_{2}^{2})^2  \right)}\nonumber\\
&&\qquad\qquad\qquad \times \frac{1}
{(q_{1} - q_{2} + p_{3})^{2} + \bigl((k_{1} - k_{2} + k_{3}')^{2}\bigl)^2}\;\;.
\end{eqnarray}
is one of the two-loop graphs which will contribute to the
fixed-point, and

\begin{eqnarray}
I_{5}\;\; =&&
\int \frac{d^{d-m}{q_{1}}d^{d-m}q_{2}d^{d-m}q_{3}d^{m}k_{1}d^{m}k_{2}
d^{m}k_{3}}
{\left( q_{1}^{2} + (k_{1}^{2})^2 \right)
\left( q_{2}^{2} + (k_{2}^{2})^2 \right)
\left( q_{3}^{2} + (k_{3}^{2})^2 \right)
[ (q_{1} + q_{2} - p)^{2} + \bigl((k_{1} + k_{2} - k')^{2}\bigr)^2]}
\nonumber\\
&&\qquad\qquad\qquad\times
\frac{1} {(q_{1} + q_{3} - p)^{2} + \bigl((k_{1} + k_{3} -
  k')^{2}\bigr)^2}
\end{eqnarray}
is the three-loop diagram contributing to the two-point vertex
function. Now we proceed to calculate these integrals using two
different approximation schemes. The philosophy to be addopted is
to simplify the calculation by making use of the homogeneity hypothesis,
as shall become clear in the following subsections.

\subsubsection{The  \lq\lq dissipative'' approximation}

As this approximation is only suited to calculate the integral as a
function of the quadratic external momenta, we set the external
momenta at the quartic directions equal to zero,
i.e. $k'=k'_1= k'_2=k'_3=0$, and $K'=k'_1+k'_2$. We shall use dimensional
regularization for the calculation of the Feynman diagrams.

Let us find out the one-loop integral $I_2$. With our choice of the
symmetry point, and introducing two Schwinger's
parameters we obtain for $I_2$:

\begin{eqnarray}
& & \int \frac{d^{d-m}{q}d^{m}k}{\left( (k^{2})^2 + (q + P)^{2}
\right) \left((k^{2})^2 + q^{2}  \right)} = \int^{\infty}_{0}\int^{\infty}_{0}
d\alpha_{1}d\alpha_{2}
\Biggl( \int d^{m}k \,\exp(-(\alpha_{1} + \alpha_{2})(k^{2})^2)
\Biggr) \nonumber \\
& & \qquad\qquad\times\int d^{d-m}q\, \exp(-(\alpha_{1} + \alpha_{2})q^{2}
- 2\alpha_{2}q.P - \alpha_{2}P^{2}) .
\end{eqnarray}
The $\vec{q}$ integral can be performed to give

\begin{eqnarray}
&& \int d^{d-m}q \,\exp(-(\alpha_{1} + \alpha_{2})q^{2} - 2\alpha_{2}q.P
- \alpha_{2}P^{2})\nonumber \\
&& \qquad\quad  = \frac{1}{2} S_{d-m} \Gamma(\frac{d-m}{2})
(\alpha_{1} + \alpha_{2})^{- \frac{d-m}{2}} \,\exp(- \frac{\alpha_{1}
\alpha_{2}P^{2}}{\alpha_{1} + \alpha_{2}})\;\;.
\end{eqnarray}
For the $\vec{k}$ integral we perform the change of variables
$r^2=k_1^2+...+k_m^2$. Now take $z=r^4$. The integral turns out to be:

\begin{equation}
\int d^{m}k \,\exp(-(\alpha_{1} + \alpha_{2})(k^{2})^2) =
\Bigl(\frac{1}{4}S_m\Bigr)\Gamma(\frac{m}{4})
 (\alpha_{1} + \alpha_{2})^{- \frac{m}{4}}.
\end{equation}
Using Eqs. (90) and (91), $I_2$ reads
\begin{eqnarray}
&& I_2= \frac{1}{2} S_{d-m}\Bigl(\frac{1}{4}S_m\Bigr)
\Gamma(\frac{d-m}{2})\Gamma(\frac{m}{4})\nonumber\\
&&\quad\times\int^{\infty}_{0}\int^{\infty}_{0}
d\alpha_{1}d\alpha_{2}\,\exp(- \frac{\alpha_{1}
\alpha_{2}P^{2}}{\alpha_{1} + \alpha_{2}})\;
(\alpha_{1} + \alpha_{2})^{- \bigl(\frac{d}{2} -\frac{m}{4}\bigr)}.
\end{eqnarray}
The remaining parametric integrals can be solved by a change of
variables followed by a rescaling \cite{AL2}. The integral is
proportional to $(P^2)^{-\frac{\epsilon_L}{2}}$.
Now we can set $P^2=\kappa^2=1$. Using the identity

\begin{equation}
\Gamma (a + b x) = \Gamma(a)\,\Bigl[\,1 + b\, x\, \psi(a) + O(x^{2})\,\Bigr],
\end{equation}
where $\psi(z) = \frac{d}{dz} ln \Gamma(z)$, one is able to
perform the $\epsilon_{L}$-expansion when the Gamma functions have non integer
arguments. Altogether, the final result for $I_2$ is:

\begin{equation}
I_{2} = \Biggl[\frac{1}{4}S_m S_{d-m}
\Gamma(2-\frac{m}{4})\Gamma(\frac{m}{4})\Biggr]\\
\frac{1}{\epsilon_{L}}\biggl(1 + [i_{2}]_m\,\epsilon_{L}\biggr)\;\;\;,
\end{equation}
\noindent where $[i_{2}]_m =  1+ \frac{1}{2}
(\psi(1) - \psi(2-\frac{m}{4}))$. The factor inside the brackets in
Eq. (94) is absorbed in a redefinition of the coupling constant.
Then the redefined integral is:

\begin{equation}
{I}_{2} = \frac{1}{\epsilon_{L}}\biggl(1 + [i_{2}]_m\,\epsilon_{L}\biggr)\;\;.
\end{equation}

Note that this expression involves no approximation. This simple result
is a consequence of the absence of the zero quartic external momenta. Had we
considered it from the beggining, we would have obtained an intermediate
integral that could not be integrated analytically. We shall discuss this
issue later in the next subsections.

However, when we go on to calculate higher loop integrals, some sort
of approximation is required, since these integrals are complicated
by the fact that even with zero external quartic momenta, the quartic
loop momenta mix up in different subdiagrams in a extremely non-trivial
form. As an example, we discuss $I_{3}$. It is given by:

\begin{equation}
I_{3}(P,K') = \int \frac{d^{d-m}{q_{1}}d^{d-m}q_{2}d^{m}k_{1}d^{m}k_{2}}
{\left( q_{1}^{2} + (k_{1}^{2})^2 \right)
\left( q_{2}^{2} + (k_{2}^{2})^2 \right)
[(q_{1} + q_{2} + P)^{2} + \bigl((k_{1} + k_{2} + K')^{2}\bigr)^2]}\;\;.
\end{equation}

Setting $K'=0$, the integral can be evaluated as outlined in \cite{AL1}.
Before making our approximation, one can choose to integrate first either
over the loop momenta $(q_{1}, k_{1})$ or over  $(q_{2}, k_{2})$.
The loop integrals to be integrated
first are referred to as the internal bubbles. By
solving the integral over $q_{2}$ first, we obtain
\begin{eqnarray}
&& I_{3}(p, 0) = \frac{1}{2} S_{d-m} \Gamma(\frac{d-m}{2})
\int \frac {d^{d-m}q_1 d^{m}k_1}{q_{1}^{2} + (k_{1}^{2})^{2}} \nonumber \\
&& \int_{0}^{\infty} \int_{0}^{\infty} d \alpha_{1} d \alpha_{2}
(\alpha_{1} + \alpha_{2})^{\frac{-(d-m)}{2}}
exp(-\frac{\alpha_{1} \alpha_{2}}{\alpha_{1} + \alpha_{2}} (q_{1} + p)^{2})
\int d^{m}k_{2} e^{-\alpha_{1} (k_{2}^{2})^{2}}
e^{-\alpha_{2} ((k_{1} + k_{2})^{2})^{2}}.
\end{eqnarray}

Now we can consider the approximation. In order to integrate over $k_{2}$,
we have to expand the argument of the last exponential. This will produce a
complicated function of $\alpha_{1}, \alpha_{2}, k_{1}$ and $k_{2}$ which
cannot be integrated analytically. Considering the remaining terms as a
damping factor to the integrand, the maximum of the integrand will be either
at $k_{1}=0$ or at $k_{1} = -2k_{2}$. The most general choice
$k_{1}= -\alpha k_{2}$ yields a hypergeometric function.
The choice $k_{1} = -2k_{2}$ implies that $k_{1}$ varies
in the internal bubble, but not arbitrarily. Its variation, however,
is dominated by $k_{2}$ through this constraint, which eliminates the
dependence on $k_{1}$ in the internal bubble. At these values of $k_{1}$,
the integration over $k_{2}$ produces a simple factor to the parametric
integral proportional to $(\alpha_{1} + \alpha_{2})^{-\frac{m}{4}}$.
This allows one to perform the remaining parametric integrals in a simple way.
After performing these integrals, they produce the factor
$((q_{1} + P)^{2})^{-\frac{\epsilon_{L}}{2}}$. Note that the diagrams
$I_{3}$ and $I_{5}$ contributing to the two-point function receive the factor
$\frac{1}{2 - \frac{m}{4}}$ after integrating over the quadratic momenta in
the external bubble. This factor will not be present in the isotropic case,
since there is no integration over quadratic momenta to be done in this case.
The resulting solution to $I_{3}(P,0)$ is a homogeneous function of the
external momenta $P$, given by:

\begin{equation}
I_{3} = - (P^{2})^{1 - \epsilon_{L}} \frac{1}{8-m}\,\frac{1}{ \epsilon_{L}}
\Biggl[1 + \biggl([i_{2}]_m + \frac{3}{4-\frac{m}{2}} + 1 \biggr)\epsilon_{L}\Biggr].
\end{equation}

The implementation of this constraint on higher-loop integrals proceeds
analogously. The constraint turns all these integrals into homogeneous
functions of the external quadratic momenta scale. One can then choose the
symmetry point as $P^{2}= \kappa_{1}^{2}= 1$, for example, in order to
define the renormalized vertices via normalization conditions.

Using this constraint we can easily find the following results at
the symmetry point \cite{AL1}:

\begin{equation}
I_{4} = \frac{1}{2 \epsilon_{L}^{2}} \Bigl(1 +
3\;[i_{2}]_m \epsilon_{L}\Bigr).
\end{equation}

The integrals $I_3^\prime$ and $I_5^\prime$ are given by

\begin{equation}
I'_{3} = - \frac{1}{8-m}\,\frac{1}{ \epsilon_{L}} \Biggl[1 +
\biggl([i_{2}]_m + \frac{3}{4-\frac{m}{2}}\biggr)\epsilon_{L}\Biggr],
\end{equation}

\begin{equation}
I'_{5} =
- \frac{1}{3\bigl(2-\frac{m}{4}\bigr)}\frac{1}{ \epsilon_{L}^{2}}
\Biggl[1 + 2\biggl([i_{2}]_m + \frac{1}{2-\frac{m}{4}}
\biggr)\epsilon_{L}\Bigr].
\end{equation}

Note that the leading singularities for $I_2$, $I_4$ are
the same as their analogous integrals in the pure $\phi^4$ theory.
However, $I_3'$ and $I_5'$ do not have the
same leading singularities for they include a factor of
$\frac{1}{(2-\frac{m}{4})}$. We then introduce a weight factor for
$I_3'$ and $I_5'$, namely $(1-\frac{m}{8})$, so that they have
the same leading singularities as in the pure $\phi^4$ theory.
The main drawback of this approximation is the failure to treat the
isotropic case. Furthermore, the introduction of weight factors to the
two and three-loop diagrams is rather undesirable. Moreover,
the constraint among the quartic loop momenta does not allow loop
momentum conservation along the quartic subspace in higher-loop diagrams.
It is then appropriate to name this approximation
``the dissipative approximation''.

It is obvious that some important detail is missing. A proper solution of
the Feynman integrals should be expressed as a homogeneous function of
both external momenta scales. We proceed to discuss the novel
approximation which presents this property.

\subsubsection{The orthogonal approximation}

Before considering the integrals to be performed, let us derive
some useful formulas which relate Gamma functions with certain
intermediate parametric integrals. They will allow us to define a
new analytic dimensional regularization procedure in the competing
subspace.

The simple integral
\begin{equation}
\int_{0}^{\infty} exp(-ax^{\mu}) dx = a^{\frac{-1}{\mu}} \frac{1}{\mu} \Gamma(\frac{1}{\mu}),
\end{equation}
can be generalized to the $m$-sphere. We shall analyse the case $\mu =2n$.
Take $r^{2} = x_{1}^{2} + ... + x_{m}^{2}$. After that take $z=r^{2n}$. Thus,
\begin{equation}
\int_{-\infty}^{\infty} dx_{1}...dx_{m} exp(-a(x_{1}^{2} + ...+x_{m}^{2})^{n})
= \frac{1}{2n} \int_{0}^{\infty} dz exp(-az) z^{\frac{m}{2n} -1} \int d\Omega_{m}.
\end{equation}
The angular integral will produce the area of the $m$-dimensional sphere,
yielding
\begin{equation}
\int_{-\infty}^{\infty} dx_{1}...dx_{m} exp(-(x_{1}^{2} + ...+x_{m}^{2})^{n})
= \frac{1}{2n} \Gamma(\frac{m}{2n}) a^{\frac{-m}{2n}} S_{m}.
\end{equation}
One can write this identity in a different way. After choosing
$r^{2}= x_{1}^{2} + ... + x_{m}^{2}$ take $y=r^{2}$ and as the
integral is given by the expression above, we obtain the intermediate
result:
\begin{equation}
\int_{0}^{\infty} dy y^{\frac{m}{2} - 1} exp(-ay^{n})
= \frac{1}{n} a^{\frac{-m}{2n}} \Gamma(\frac{m}{2n}).
\end{equation}
Henceafter we shall keep $n=2$. The following step is to calculate the
integral
\begin{equation}
\int_{-\infty}^{\infty} exp(-ax^{4} - bx^{2}) dx = 2 \int_{0}^{\infty}
exp(-ax^{4} - bx^{2}) dx.
\end{equation}
The exact answer is given in terms of a Bessel function of a certain
combination of $a$ and $b$. We wish to pick out only the piece which yields
the correct homogenous function of $a$, i.e., only one term of the series.
This can be achieved as follows. Choose $y=x^{2}$. One obtains
\begin{equation}
\int_{-\infty}^{\infty} exp(-ax^{4} - bx^{2}) dx = exp{(\frac{b^{2}}{4a})}
\int_{0}^{\infty} exp(-a(y +\frac{b}{2a})^{2}) y^{\frac{-1}{2}}dy.
\end{equation}
We then choose $y' = y + \frac{b}{2a}$ implying that
\begin{eqnarray}
&\int_{-\infty}^{\infty} exp(-ax^{4} - bx^{2}) dx =
exp[\frac{b^{2}}{4a}] [\int_{0}^{\infty} exp(-ay'^{2}) (y' -\frac{b}{2a})^{\frac{-1}{2}}dy' \nonumber\\
& - \int_{0}^{\frac{b}{2a}} exp(-ay'^{2}) (y' -\frac{b}{2a})^{\frac{-1}{2}}dy'].
\end{eqnarray}
Since we are dealing with convergent integrals, we can perform the
approximation $(y'-\frac{b}{2a})^{\frac{-1}{2}} = y'^{-\frac{1}{2}} + ...$,
and the remaining terms will be subtracted from the last integral, which is
a sort of error function. The original integral is then approximated by its
leading contribution
\begin{equation}
\int_{-\infty}^{\infty} exp(-ax^{4} - bx^{2}) dx \cong
exp{(\frac{b^{2}}{4a})} \int_{0}^{\infty} exp(-ay'^{2})
y'^{-\frac{1}{2}}dy'= exp{(\frac{b^{2}}{4a})} \frac{1}{2}
\Gamma(\frac{1}{4}) a^{-\frac{1}{4}}.
\end{equation}
It can be shown in a straightforward way that for the $m$-sphere this result
generalizes to
\begin{eqnarray}
& \int_{-\infty}^{\infty} exp[-a(x_{1}^{2} + ... + x_{m}^{2})^{2}
- b(x_{1}^{2} + ... + x_{m}^{2})] dx_{1}...dx_{m}  \cong
exp{(\frac{b^{2}}{4a})} S_{m} \times \nonumber\\
& \int_{0}^{\infty} exp(-ay'^{2}) y'^{\frac{m}{2} -1}dy'=
exp{(\frac{b^{2}}{4a})} \frac{1}{4} S_{m} \Gamma(\frac{m}{4})
a^{-\frac{m}{4}}.
\end{eqnarray}
We now focus our attention in the integral
\begin{equation}
\int_{-\infty}^{\infty} \frac{dx_{1}...dx_{m}}
{[(x_{1}^{2} + ...+x_{m}^{2})^{2} + 2a(x_{1}^{2} + ...+x_{m}^{2}) + m^{2}]
^{\beta}}.
\end{equation}
Take $r^{2} = x_{1}^{2} + ...+x_{m}^{2}$. Make the change of variables in
the radial coordinate $z=r^{2}$. After that take $z' = z + a$. We then obtain
\begin{eqnarray}
& \int \frac{d^{m}x}{[(x_{1}^{2} + ...+x_{m}^{2})^{2} + 2a(x_{1}^{2} + ...+x_{m}^{2}) + m^{2}]^{\beta}} = \frac{1}{2} S_{m} \nonumber\\
& [\int_{0}^{\infty} \frac{(z'-a)^{\frac{m}{2} -1} dz'}
{(z'^{2} + m'^{2})^{\beta}} - \int_{0}^{a} \frac {(z'-a)^{\frac{m}{2} -1} dz'}
{(z'^{2} + m'^{2})^{\beta}}],
\end{eqnarray}
where $m'^{2} = m^{2} + a^{2}$. Taking $z''=z'^{2}$, expanding the numerator
in the first integral, i.e., keeping only the leading term and getting rid of
the infinite terms to be subtracted from the second integral, one can
write this integral in the approximated form
\begin{eqnarray}
& \int \frac{d^{m}x}{[(x_{1}^{2} + ...+x_{m}^{2})^{2} + 2a(x_{1}^{2} + ...+x_{m}^{2}) + m^{2}]^{\beta}} \cong \frac{1}{4} S_{m}
(m^{2} - a^{2})^{-\beta + \frac{m}{4}} \nonumber\\
& \frac{\Gamma(\frac{m}{4}) \Gamma(\beta -\frac{m}{4})}{\Gamma(\beta)}.
\end{eqnarray}
We have all ingredients to perform Feynman integrals for arbitrary $m$.
We start by considering the simplest integral, the one-loop integral
contributing to the coupling constant, that is,

\begin{equation}
I_2 =  \int \frac{d^{d-m}q d^{m}k}{[\bigl((k + K^{'})^{2}\bigr)^2 +
(q + P)^{2}] \left( (k^{2})^2 + q^{2}  \right)}\;\;\;.
\end{equation}
We can use two Schwinger parameters and integrate over the quadratic
momenta. Using the formula derived above
\begin{equation}
\int \exp(-p^{2}) d^{d}q = \frac{1}{2} S_{d} \Gamma(\frac{d}{2}),
\end{equation}
we obtain
\begin{eqnarray}
&& I_2= \frac{1}{2} S_{d-m}
\Gamma(\frac{d-m}{2}) \int^{\infty}_{0}\int^{\infty}_{0}
d\alpha_{1}d\alpha_{2}\,\exp(- \frac{\alpha_{1}
\alpha_{2}P^{2}}{\alpha_{1} + \alpha_{2}})\nonumber\\
&& \quad\times(\alpha_{1} + \alpha_{2})^{- \bigl(\frac{d-m}{2} \bigr)}
\int d^{m}k \exp(-\alpha_{1}(k^{2})^{2} - \alpha_{2}((k + K^{'})^{2})^{2}).
\end{eqnarray}
We can now expand the argument of the last exponential. This integral
cannot be performed analytically. We are interested in
the solution of this integral in a form that preserves
homogeneity in both external momenta. Some simplifying condition
should be tried to achieve this goal.

The most general approximation to calculate
this type of integral which is homogeneous in the external momenta scales
can be understood as follows. In the first place, if we set $k.K^{'} = 0$
inside the integral, that has the virtue of eliminating odd powers of the
quartic external momenta. Thus, the integral becomes:
\begin{eqnarray}
& \int d^{m}k \exp(-\alpha_{1}(k^{2})^{2} - \alpha_{2}((k + K^{'})^{2})^{2}) =
\int d^{m}k \nonumber\\
& \exp(-(\alpha_{1} + \alpha_{2})(k^{2})^{2}
- 2 \alpha_{2} k^{2} (K')^{2} - \alpha_{2}((K^{'})^{2})^{2}).
\end{eqnarray}
Using Eq.(111), we have for the last quartic
momenta integral:
\begin{eqnarray}
& \int d^{m}k \exp(-\alpha_{1}(k^{2})^{2} - \alpha_{2}((k + K^{'})^{2})^{2}) =
S_{m} \frac{1}{4} \Gamma(\frac{m}{4})
(\alpha_{1} + \alpha_{2})^{-\frac{m}{4}}\nonumber\\
& \times \; \exp(- \frac{\alpha_{1} \alpha_{2}  ((K^{'})^{2})^{2}}{\alpha_{1} + \alpha_{2}}).
\end{eqnarray}
We can then express the integral in the following form:
\begin{eqnarray}
&& I_2= \frac{1}{8} S_{d-m} S_{m}
\Gamma(\frac{d-m}{2})\Gamma(\frac{m}{4})\nonumber\\
&&\quad\times\int^{\infty}_{0}\int^{\infty}_{0}
d\alpha_{1}d\alpha_{2}\,\exp(- \frac{\alpha_{1}
\alpha_{2}(P^{2} + ((K')^{2})^{2}}{\alpha_{1} + \alpha_{2}})\;
(\alpha_{1} + \alpha_{2})^{- \bigl(\frac{d}{2} -\frac{m}{4}\bigr)}.
\end{eqnarray}
Take $x=\alpha_{1} (P^{2} + (K'^{2})^{2})$ and
$y = \alpha_{2}(P^{2} + (K'^{2})^{2})$. After that, define
$v = \frac{x}{x+y}$. Thus, the parametric integrals can be done easily by this change of variables. Then, use the identity
\begin{equation}
\Gamma (a + b x) = \Gamma(a)\,\Bigl[\,1 + b\, x\, \psi(a) + O(x^{2})\,\Bigr],
\end{equation}
where $\psi(z) = \frac{d}{dz} ln \Gamma(z)$. This will result in the
following expression for $I_{2}$:
\begin{eqnarray}
& I_{2} = \frac{1}{2}[\frac{1}{4} S_{(d-m)}S_{m} \Gamma(2 - \frac{m}{4})
\Gamma(\frac{m}{4})] (1 - \frac{\epsilon_{L}}{2} \psi(2 - \frac{m}{4}))
\Gamma(\frac{\epsilon_{L}}{2})\nonumber\\
& \times \; \int_{0}^{1} dv(v(1-v)(P^{2} + ((K')^{2})^{2}))^{\frac{-\epsilon_{L}}{2}}.
\end{eqnarray}
This is a homogeneous function (with the same homogeneity degree)
in $(P,K')$ just as advertised. But this is not the answer yet. The
factor $[\frac{1}{4} S_{(d-m)}S_{\frac{m}{2}} \Gamma(2 - \frac{m}{4})
\Gamma(\frac{m}{4})]$ can be absorbed in a redefinition of the
coupling constant. Hence, we shall absorb exactly this factor after
performing each loop integral. Furthermore, the last integral can be
expanded as
\begin{equation}
\int_{0}^{1} dv (v(1-v)(P^{2} + ((K')^{2})^{2}))^{\frac{-\epsilon_{L}}{2}} =
1 - \frac{\epsilon_{L}}{2}L(P,K'),
\end{equation}
where
\begin{equation}
L(P,K') =
\int_{0}^{1} dv   \;\;ln[v(1-v)(P^{2} + ((K')^{2})^{2})].
\end{equation}
Thus, we find the following result for this integral:
\begin{equation}
{I}_{2}(P,K') =
\frac{1}{\epsilon_{L}}\biggl(1 + ([i_{2}]_{m} - 1) \epsilon_{L}
-\frac{\epsilon_{L}}{2} L(P,K') \biggr)\;\;.
\end{equation}
This is the form suitable for renormalizing using minimal subtraction.
On the other hand, for normalization conditions one has:
\begin{equation}
{I}_{2 SP_{1}} = {I}_{2 SP_{2}} =
\frac{1}{\epsilon_{L}}\biggl(1 + [i_{2}]_{m}\,\epsilon_{L} \biggr)\;\;,
\end{equation}
since $L(SP_{1}=SP_{2})=-2$, with $SP_{1}\equiv (P^{2}=1,K'=0)$ and
$SP_{2} \equiv (P=0,(K')^{2}=1)$.
When we calculated $I_{2}(P,K'=0)$ in the last subsection, the orthogonality
condition $k.K'=0$ between the loop momenta and the external momenta
along the quartic subspace was trivial. In the calculation of $I_{2}(P,K')$,
the orthogonality condition allowed the solution to this integral with
the correct homogeneous properties in both external momenta scales.

We can now turn our attention to the higher-loop integrals. The
simplifying condition $k.K'=0$ for the one-loop integral can be easily
generalized to the higher-loop integrals by stating that {\it the loop momenta in a given bubble (subdiagram) is orthogonal to all loop momenta not
belonging to that bubble}. Let us see how this works in the
calculation of the ``sunset'' two-loop integral $I_{3}$ contributing to
the two-point function, given by the following expression:

\begin{equation}
I_{3}(P,K') = \int \frac{d^{d-m}{q_{1}}d^{d-m}q_{2}d^{m}k_{1}d^{m}k_{2}}
{\left( q_{1}^{2} + (k_{1}^{2})^2 \right)
\left( q_{2}^{2} + (k_{2}^{2})^2 \right)
[(q_{1} + q_{2} + P)^{2} + \bigl((k_{1} + k_{2} + K')^{2}\bigr)^2]}\;\;.
\end{equation}

We can choose to integrate first over the loop momenta $(q_{1},k_{1})$
or over $(q_{2},k_{2})$. The loop integrals to be integrated first are
referred to as the internal bubbles.  By
solving the integral over $q_{2}$ first, we obtain
\begin{eqnarray}
&& I_{3}(P, K') = \frac{1}{2} S_{d-m} \Gamma(\frac{d-m}{2})
\int \frac {d^{d-m}q_1 d^{m}k_1}{q_{1}^{2} + (k_{1}^{2})^{2}}
\int_{0}^{\infty} \int_{0}^{\infty} d \alpha_{1} d \alpha_{2}
(\alpha_{1} + \alpha_{2})^{\frac{-(d-m)}{2}} \nonumber \\
&& exp(-\frac{\alpha_{1} \alpha_{2}}{\alpha_{1} + \alpha_{2}} (q_{1} + P)^{2})
\int d^{m}k_{2} e^{-\alpha_{1} (k_{2}^{2})^{2}}
e^{-\alpha_{2} ((k_{1} + k_{2} + K')^{2})^{2}}.
\end{eqnarray}
We define $K''= k_{1} + K'$ into the argument of the last exponential,
and integrate over $k_{2}$ using the condition $k_{2}.K''=0$.
Make the change of variables $k_{2}^{2} = p$ and integrate over
$k_{2}$ (or $p$). Using Eq.(111) we find:
\begin{eqnarray}
&& I_{3}(P, K') = \frac{1}{8} S_{d-m} S_{m} \Gamma(\frac{d-m}{2})
\Gamma(\frac{m}{4})
\int \frac {d^{d-m}q_1 d^{m}k_1}{q_{1}^{2} + (k_{1}^{2})^{2}} \nonumber \\
&& \int_{0}^{\infty} \int_{0}^{\infty} d \alpha_{1} d \alpha_{2}
(\alpha_{1} + \alpha_{2})^{-[\frac{(d-m)}{2} + \frac{m}{4}]}
exp(-\frac{\alpha_{1} \alpha_{2}}{\alpha_{1} + \alpha_{2}}
[(q_{1} + P)^{2} + ((k_{1} + K')^{2})^{2}]).
\end{eqnarray}
The parametric integrals can be solved as before and we have:
\begin{equation}
I_{3}(P, K') = \frac{1}{8} S_{d-m} S_{m} \Gamma(\frac{d-m}{2})
\Gamma(\frac{m}{4})
\int \frac {d^{d-m}q_1 d^{m}k_1}{(q_{1}^{2} + (k_{1}^{2})^{2})[(q_{1} + P)^{2} + ((k_{1} + K')^{2})^{2}]^{\frac{\epsilon_{L}}{2}}}.
\end{equation}
We can now use Eq.(94) and absorbing the angular geometric factor for
the first loop integral we obtain:
\begin{equation}
I_{3}(P, K') = \frac{1}{\epsilon_{L}}(1+[i_{2}]_{m} \epsilon_{L})
\int \frac {d^{d-m}q_{1} d^{m}k_{1}}{(q_{1}^{2} + (k_{1}^{2})^{2})[(q_{1} + P)^{2} + ((k_{1} + K')^{2})^{2}]^{\frac{\epsilon_{L}}{2}}}.
\end{equation}
Let $i_{3}(P,K')$ be the last integral above. In the remainder, we employ
a Feynman parameter. The integral $i_{3}(P,K')$ can be expressed in the
following form:

\begin{eqnarray}
&& i_{3}(P, K') = \frac{\Gamma(1 + \frac{\epsilon_{L}}{2})}{\Gamma(\frac{\epsilon_{L}}{2})}\int_{0}^{1} dx x^{\frac{\epsilon_{L}}{2} -1}\nonumber\\
&&\int\frac {d^{d-m}q_{1}d^{m}k_{1}}{(q_{1}^{2} + 2x P.q_{1} + xP^{2} + (1-x) (k_{1}^{2})^{2} + x((k_{1} + K')^{2})^{2})^{1 + \frac{\epsilon_{L}}{2}}}.
\end{eqnarray}
After that, take  the orthogonality condition $k_{1}.K'=0$.
In order to solve the integral over the quadratic momenta $q_{1}$
we shall make use of the relation:
\begin{equation}
\int \frac {d^{d-m}q_1}{(q_{1}^{2} + 2 k.q_{1} + m^{2})^{\alpha}} =
\frac{1}{2} \frac{\Gamma(\frac{(d-m)}{2}) \Gamma(\alpha - \frac{(d-m)}{2}) (m^{2} - k^{2})^{\frac{(d-m)}{2} - \alpha}}{\Gamma(\alpha)} S_{d-m}.
\end{equation}
Thus, we obtain
\begin{eqnarray}
&& i_{3}(P, K') = \frac{S_{d-m}}{2} \frac{\Gamma(2 - \frac{m}{4} + \epsilon_{L})
\Gamma(-1 + \frac{m}{4} + \epsilon_{L})}{\Gamma(\frac{\epsilon_{L}}{2})}\int_{0}^{1} dx x^{\frac{\epsilon_{L}}{2} -1}\nonumber\\
&&\int\frac {d^{m}k_{1}}{((k_{1}^{2})^{2} + P^{2} x(1-x) +
x(2 k_{1}^{2}K'^{2} + (K'^{2})^{2}))^{-1 + \frac{m}{4} + \epsilon_{L}}}.
\end{eqnarray}
Now, using Eq.(114), we can integrate over the quartic momenta
$k_{1}$ obtaining:
\begin{eqnarray}
&& i_{3}(P, K') = \frac{S_{d-m} S_{m}}{8} \frac{\Gamma(2 - \frac{m}{4} +
\frac{\epsilon_{L}}{2}) \Gamma(\frac{m}{4})
\Gamma(-1 + \epsilon_{L})}{\Gamma(\frac{\epsilon_{L}}{2})} \nonumber\\
&& \times \int_{0}^{1} dx x^{\frac{\epsilon_{L}}{2} -1} [x(1-x)(P^{2} + (K'^{2})^{2})]^{1 - \epsilon_{L}}.
\end{eqnarray}
Expanding the resulting $\Gamma$ functions and  absorbing the geometric
angular factor discussed above,
we find:
\begin{equation}
I_{3}(P, K') = (P^{2} + (K'^{2})^{2})\frac{-1}{8 \epsilon_{L}}
(1+2[i_{2}]_{m} \epsilon_{L} -\frac{3}{4} \epsilon_{L} -
2 \epsilon_{L} L_{3}(P, K')),
\end{equation}
where
\begin{equation}
L_{3}(P, K') = \int_{0}^{1} dx (1-x) ln[(P^{2} + (K'^{2})^{2}) x(1-x)].
\end{equation}
At the symmetry points $SP_{1}$ or $SP_{2}$, it can be rewritten as
\begin{equation}
I_{3 SP_{1}} = I_{3 SP_{2}} = \frac{-1}{8 \epsilon_{L}}
(1+2[i_{2}]_{m} \epsilon_{L} + \frac{5}{4} \epsilon_{L}).
\end{equation}
From the above equation we can derive the expressions:
\begin{equation}
I_{3 SP_{1}}^{'} (\equiv \frac{\partial I_{3 SP_{1}}}{\partial P^{2}})=
I_{3 SP_{2}}^{'} (\equiv \frac{\partial I_{3 SP_{2}}}{\partial K'^{4}}) =
\frac{-1}{8 \epsilon_{L}}
(1+2[i_{2}]_{m} \epsilon_{L} + \frac{1}{4} \epsilon_{L}).
\end{equation}

Let us now proceed to discuss the other required loop integrals. Consider
\begin{eqnarray}
I_{5}\;\; =&&
\int \frac{d^{d-m}{q_{1}}d^{d-m}q_{2}d^{d-m}q_{3}d^{m}k_{1}d^{m}k_{2}
d^{m}k_{3}}
{\left( q_{1}^{2} + (k_{1}^{2})^2 \right)
\left( q_{2}^{2} + (k_{2}^{2})^2 \right)
\left( q_{3}^{2} + (k_{3}^{2})^2 \right)
[ (q_{1} + q_{2} - P)^{2} + \bigl((k_{1} + k_{2} - K')^{2}\bigr)^2]}
\nonumber\\
&&\qquad\qquad\qquad\times
\frac{1} {(q_{1} + q_{3} - P)^{2} + \bigl((k_{1} + k_{3} -
  K')^{2}\bigr)^2},
\end{eqnarray}
which is the three-loop diagram contributing to the two-point vertex
function. This integral is symmetric under a change in the dummy loop
momenta  $q_{2} \rightarrow q_{3}$ and $k_{2} \rightarrow k_{3}$.
Let us analyse the integrations over $q_{2}, k_{2}$ and $q_{3}, k_{3}$. We
use the condition $k_{2}.(k_{1} - K')=0$ when integrating over $k_{2}$
as well as $k_{3}.(k_{1} - K')=0$ when performing the integral over
$k_{3}$. The two internal bubbles, which are represented by the
integrals over $(q_{2}, k_{2})$ and $(q_{3}, k_{3})$, respectively,
give actually the same result, namely $I_{2}((q_{1} - P), (k_{1} - K'))$.
Next take $P\rightarrow -P$, $K'\rightarrow -K'$. Therefore, we obtain the
following expression:
\begin{equation}
I_{5 }(P, K') = \frac{1}{\epsilon_{L}^{2}}(1 + 2 [i_{2}]_{m} \epsilon_{L})
\int \frac {d^{d-m}q_{1}d^{m}k_{1}}{(q_{1}^{2} + (k_{1}^{2})^{2})[(q_{1} + P)^{2} + ((k_{1} + K')^{2})^{2}]^{\epsilon_{L}}}.
\end{equation}
We employ a Feynman parameter in analogy to what was done in the calculation
of $I_{3}$ and working out the details we find:
\begin{equation}
I_{5}(P, K') = (P^{2} + (K'^{2})^{2})\frac{-1}{6 \epsilon_{L}^{2}}
(1+3[i_{2}]_{m} \epsilon_{L} - \epsilon_{L} - 3 \epsilon_{L} L_{3}(P, K')),
\end{equation}
At the symmetry points $SP_{1}$, $SP_{2}$ we find:
\begin{equation}
I_{5 SP_{1}}^{'} (\equiv \frac{\partial I_{5 SP_{1}}}{\partial P^{2}})=
I_{5 SP_{2}}^{'} (\equiv \frac{\partial I_{5 SP_{2}}}{\partial K'^{4}}) =
\frac{-1}{6 \epsilon_{L}^{2}}
(1+3[i_{2}]_{m} \epsilon_{L} + \frac{1}{2} \epsilon_{L}).
\end{equation}

We are left with the task of calculating one of the two-loop diagrams
contributing to the four-point function
\begin{eqnarray}
I_{4}\;\; =&& \int \frac{d^{d-m}{q_{1}}d^{d-m}q_{2}d^{m}k_{1}d^{m}k_{2}}
{\left( q_{1}^{2} + (k_{1}^{2})^2 \right)
\left( (P - q_{1})^{2} + \bigl((K' - k_{1})^{2}\bigr)^2  \right)
\left( q_{2}^{2} + (k_{2}^{2})^2  \right)}\nonumber\\
&&\qquad\qquad\qquad \times \frac{1}
{(q_{1} - q_{2} + p_{3})^{2} + \bigl((k_{1} - k_{2} + k_{3}')^{2}\bigl)^2}\;\;.
\end{eqnarray}
Notice that $P= p_{1} + p_{2}$, $p_{i}$ ($i=1,...,3$) are external
momenta perpendicular to the competing axes whereas $K'= k'_{1} + k'_{2}$,
and $k'_{i}$ ($i=1,...,3$) are the external momenta along the competition
directions. We can integrate first over the bubble $(q_{2}, k_{2})$. Using
Schwinger parameters, and absorbing the geometric angular factor for the
first bubble we obtain:

\begin{eqnarray}
I_{4} = && \frac{1}{\epsilon_{L}}(1 + [i_{2}]_{m} \epsilon_{L})\int \frac{d^{d-m}{q_{1}}d^{m}k_{1}}
{\left( q_{1}^{2} + (k_{1}^{2})^2 \right)
\left( (P - q_{1})^{2} +  ((K' - k_{1})^{2})^2  \right)}\nonumber\\
&& \times \; \; \frac{1}{[(q_{1}+p_{3})^2 + ((k_{1} + k'_{3})^{2})^{2}]^{\frac{\epsilon_L}{2}}}\;\;.
\end{eqnarray}
Using a Feynman parameter one can write this in the form
\begin{eqnarray}
I_{4} = &&f_{m}(\epsilon_{L}) \,\int_0^1 dz\,\int
\frac{d^{d-m} q_{1} d^{m}k_{1}}
{[ q_{1}^{2} - 2z\,P.q_1+zP^2 + (k_{1}^{2})^{2} + z((K'^{2})^{2} +
2K'^{2} k_{1}^{2})]^{2}}\nonumber\\
&&\times \frac{1}{[(q_1+p_3)^2 + ((k_{1} + k'_{3})^{2})^{2} ]^{\frac{\epsilon_L}{2}}} \;,
\end{eqnarray}
where we defined the quantity
$f_{m}(\epsilon_{L})=\frac{1}{\epsilon_{L}}(1 + [i_{2}]_{m} \epsilon_{L})$,
which is the one-loop subdiagram with the angular factor already absorbed.
Using another Feynman parameter to fold the two denominators in the
last expression, integrating over $p_{1}, k_{1}$ (recalling the
orthogonality condition already stated), the integral turns out to be

\begin{eqnarray}
& I_{4} = \frac{1}{8} f_{m}(\epsilon_{L})\,
\frac{\Gamma(\epsilon_L) \Gamma(\frac{m}{4}) \Gamma(2 - \frac{m}{4} -\frac{\epsilon_{L}}{2})}{\Gamma\biggl(\frac{\epsilon_L}{2}\biggr)}
\,\int_0^1 dy\, y\,(1-y)^{\frac{1}{2}\epsilon_L-1} \int_0^1 dz
\biggl[ yz(1-yz)(P^{2} + (K'^{2})^{2})\nonumber\\
& +y(1-y)(p_{3}^{2} + ((k'_{3})^{2})^{2})
+2yz(1-y)(p_3.P+ (k'_{3})^{2} (K')^{2})\biggr]^{-\epsilon_L} S_{d-m} S_{m}.
\end{eqnarray}
The integral  over $y$ is singular at $y=1$ when
$\epsilon_L=0$. We only need to replace the value $y=1$ inside the
integral over $z$ \cite{amit}, and integrate over $y$
afterwards, obtaining after the absorption of the geometric factor

\begin{equation}
I_{4} = \frac{1}{2 \epsilon_{L}^{2}} \Bigl(1 +
2\;[i_{2}]_m \epsilon_{L} -\frac{3}{2} \epsilon_{L} - \epsilon_{L} L(P,K')\Bigr).
\end{equation}

This is the most appropriate form to carry out the renormalization using
minimal subtraction. In terms of normalization conditions, we find the
value of this integral at the symmetry points discussed before:

\begin{equation}
I_{4 SP_{1}} = I_{4 SP_{2}} = \frac{1}{2 \epsilon_{L}^{2}} \Bigl(1 +
2\;[i_{2}]_m \epsilon_{L} +\frac{1}{2} \epsilon_{L}\Bigr).
\end{equation}
Thus, we have successfully devised a new regularization procedure to
calculate Feynman integrals whose propagators have quartic powers of
momenta. It is tempting to define the measure of the $m$-dimensional
sphere in terms of a half integer measure. In fact, taking $k=p^{2n}$,
one has
$d^{m}k \equiv d^{\frac{m}{2n}}p = \frac {1}{2n} p^{\frac{m}{2n}-1}dp
d\Omega_{m}$.
Hence, the approximation ammounts to take the new ``measure''
$d^{\frac{m}{2n}}p$ invariant under translations $p'=p+a$.

Note that this approximation is much better than the former dissipative
approximation for the following reasons. First, we do not have to
introduce any diagram factor, since after absorbing the geometric angular
factor the leading singularities in $\epsilon_{L}$ are the same as those
in the standard $\phi^{4}$ field theory present in $\epsilon$ ($m=0$). Second,
we have an expression in terms of arbitrary external momenta, which permits
the computation of all the critical exponents in a completely independent
manner using renormalization group transformations either perpendicular
or parallel to the competition axes. Third, this can be easily adapted to the
isotropic behavior. The three weak points of the dissipative approximation
have been overcome in the orthogonal approximation.

\subsection{Isotropic}

The new orthogonal approximation can now be used to obtain the solutions
to the Feynman integrals in the isotropic case. At the Lifshitz point
$\delta_{0} = 0$ all the quadratic momenta disappear and only quartic
momenta are defined.

Consider the integral
\begin{equation}
I_2 =  \int \frac{d^{m}k}{\bigl((k + K^{'})^{2}\bigr)^2  (k^{2})^2 }\;\;\;.
\end{equation}
It is the isotropic counterpart of the one-loop integral contributing
to the four-point function. Using two Schwinger parameters and the
orthogonality condition $k.K'=0$, we find:
\begin{equation}
I_{2}(K') = \int d^{m}k \int_{0}^{\infty} \int_{0}^{\infty}
d \alpha_{1} d \alpha_{2} e^{-(\alpha_{1} + \alpha_{2})(k_{2}^{2})^{2}}
e^{-2\alpha_{2} K'^{2} k^{2}} e^{-\alpha_{2}(K'^{2})^{2}}.
\end{equation}
Now performing the transformation $k^{2}=p$ the volume element transforms
to $d^{m}k=\frac{1}{2} p^{\frac{m}{2} -1}dp d\Omega_{m} \equiv d^{\frac{m}{2}}p$. The
former quartic integral turns into a quadratic integral over $p$. After
neglecting the infinite terms which change the measure $d^{\frac{m}{2}}k$
under the translation $y'=y+\frac{b}{2a}$, only the leading contribution
is sorted out and we have
\begin{equation}
\int d^{m}k e^{-a (k^{2})^{2} -b k^{2}} = \int d^{\frac{m}{2}}p e^{-a p^{2} -b p} \cong a^{-\frac{m}{4}} e^{\frac{b^{2}}{4a}} \frac{1}{4} \Gamma(\frac{m}{4})
S_{m} .
\end{equation}

Replacing this result into the expression of $I_{2}(K')$, one finds
\begin{equation}
I_{2}(K') = \frac{1}{4} \Gamma(\frac{m}{4}) S_{m}
\int_{0}^{\infty} \int_{0}^{\infty} d \alpha_{1} d \alpha_{2}
(\alpha_{1} + \alpha_{2})^{-\frac{m}{4}}
\exp(-\frac{\alpha_{1} \alpha_{2} (K'^{2})^{2}}{\alpha_{1} + \alpha_{2}}).
\end{equation}
Now, use a change of variables and a rescaling to realize the remaining
parametric integrals analogously to what was done in the anisotropic case.
Making the continuation  $m = 8 - \epsilon_{L}$, the integral can be
expressed in the following intermediate step:
\begin{equation}
I_{2}(K') = \frac{1}{4} \Gamma(2 - \frac{\epsilon_{L}}{4})
\Gamma(\frac{\epsilon_{L}}{4}) S_{m}
\int_{0}^{1} dv [v(1-v) (K'^{2})^{2}]^{-\frac{\epsilon_{L}}{4}},
\end{equation}
and the integration over $v$ produces the result
\begin{equation}
I_{2}(K') = \frac{S_{m}}{\epsilon_{L}} [ 1 - \frac{\epsilon_{L}}{4} (1+L(K'^{2}))].
\end{equation}
We absorb the factor of $S_{m}$ in this integral through a
redefinition of the coupling constant. Henceafter we shall absorb this
factor when calculating each loop integral in analogy to the discussion for
the anisotropic behavior. Note that this absorption factor is different from
the one appearing for the anisotropic case in the limit
$d \rightarrow m =8 -\epsilon_{L}$. In the anisotropic case the geometric
angular factor becomes singular in the above isotropic limit indicating the
failure of this attempt of extrapolating from one case to another. This is
a more compelling technical reason for the statement that the isotropic
and anisotropic cases are completely distinct. Therefore,
\begin{equation}
I_{2}(K') = \frac{1}{\epsilon_{L}} [ 1 - \frac{\epsilon_{L}}{4} (1+L(K'^{2}))].
\end{equation}
In terms of a symmetry point $(K'^{2})^{2} = 1$ convenient whenever
normalization conditions are used, we obtain
\begin{equation}
I_{2}(K') = \frac{1}{\epsilon_{L}} [ 1 +\frac{\epsilon_{L}}{4}].
\end{equation}

We can go on to evaluate the other required higher loop
integrals. The systematics is the same: solve the subdiagrams
using the intermediate step Eq.(153) and then use Feynman parameters
to solve the parametric integrals left over.

Let us calculate
\begin{equation}
I_{3} = \int \frac{d^{m}k_{1} d^{m}k_{2}}{\bigl((k_{1} + k_{2} + K^{'})^{2}\bigr)^2  (k_{1}^{2})^2  (k_{2}^{2})^2}\;\;\;,
\end{equation}
the two-loop \lq\lq sunset'' Feynman diagram of the two-point
function in the isotropic case. Integrate first over $k_{2}$.
Take $K''= k_{1} + K'$ and use the condition $k_{2}.K'' = 0$ to obtain:
\begin{equation}
I_{3} = \frac{1}{\epsilon_{L}}[1 + \frac{\epsilon_{L}}{4}]
\int \frac{d^{m}k_{1}}
{[\bigl((k_{1} + K^{'})^{2}\bigr)^2]^{\frac{\epsilon_{L}}{4}}  (k_{1}^{2})^{2}}\;.
\end{equation}
Now using a Feynman parameter, integrating over $k_{1}$,
taking $m=8-\epsilon_{L}$, and employing the formula
\begin{equation}
\int \frac {d^{\frac{m}{2}}q}{(q^{2} + 2 k.q + m^{2})^{\alpha}} \cong
\frac{1}{4} \frac{\Gamma(\frac{m}{4}) \Gamma(\alpha - \frac{m}{4})
(m^{2} - k^{2})^{\frac{m}{4} - \alpha} S_{m}}{\Gamma(\alpha)},
\end{equation}
one can express $I_{3}$ as
\begin{equation}
I_{3} = \frac{S_{m} \Gamma(2 - \frac{\epsilon_{L}}{4}) \Gamma(-1 + \frac{\epsilon_{L}}{2})}{4 \epsilon_{L} \Gamma(\frac{\epsilon_{L}}{4})}
[1 + \frac{\epsilon_{L}}{4}] \int_{0}^{1} dx
[(K'^{2})^{2} x(1-x)]^{1-\frac{\epsilon_{L}}{2}}
x^{-1+\frac{\epsilon_{L}}{4}} .
\end{equation}
We can rewrite this expression in terms of the integral $L_{3}(K'^{2})$
defined in the last section in the form
\begin{equation}
I_{3} = \frac{S_{m} \Gamma(2 - \frac{\epsilon_{L}}{4}) \Gamma(-1 + \frac{\epsilon_{L}}{2})}{4 \epsilon_{L} \Gamma(\frac{\epsilon_{L}}{4})}
[1 + \frac{\epsilon_{L}}{4}] [\frac{1}{2} -\frac{3 \epsilon_{L}}{16}
- \frac{\epsilon_{L}}{2} L_{3}(K'^{2})].
\end{equation}

Expanding the Gamma functions and absorbing $S_{m}$, it is easy
to show that
\begin{equation}
I_{3} = -\frac{(K'^{2})^{2}}{16 \epsilon_{L}}[1 + \epsilon_{L}(\frac{1}{8}
- L_{3}(K'^{2}))].
\end{equation}
At the symmetry point, this can be expressed as:
\begin{equation}
I_{3} = -\frac{1}{16 \epsilon_{L}}[1 + \frac{9}{8} \epsilon_{L}],
\end{equation}
leading to
\begin{equation}
\frac{\partial I_{3}}{\partial (K'^{2})^{2}}|_{SP_{3}} = I_{3}^{'} =
-\frac{1}{16 \epsilon_{L}}[1 + \frac{5}{8} \epsilon_{L}].
\end{equation}
We can carry out the calculation of the other integrals using the same
reasoning. The 3-loop integral $I_{5}$ is given by
\begin{equation}
I_{5} = \int \frac{d^{m}k_{1} d^{m}k_{2}d^{m}k_{3}}
{\bigl((k_{1} + k_{2} + K^{'})^{2}\bigr)^{2} \bigl((k_{1} + k_{3} + K^{'})^{2}\bigr)^{2} (k_{1}^{2})^{2}  (k_{2}^{2})^{2}  (k_{3}^{2})^{2} }\;\;\;,
\end{equation}
where we took for convenience $K'\rightarrow -K'$. The integrals over
$k_{2}$ and $k_{3}$ are identical. Hence
\begin{equation}
I_{5} = \frac{1}{\epsilon_{L}^{2}}[1 + \frac{\epsilon_{L}}{2}]
\int \frac{d^{m}k_{1}}
{[\bigl((k_{1} + K^{'})^{2}\bigr)^2]^{\frac{\epsilon_{L}}{2}}  (k_{1}^{2})^{2}}\;.
\end{equation}
Employing the same techniques as in the calculation of $I_{3}$ we obtain
\begin{equation}
I_{5} = -\frac{(K'^{2})^{2}}{12 \epsilon_{L}^{2}}[1 + \epsilon_{L}(\frac{1}{4}
- \frac{3}{2}L_{3}(K'^{2}))].
\end{equation}
At the symmetry point, the derivative of $I_{5}$ with respect to the
external momenta is given by
\begin{equation}
\frac{\partial I_{5}}{\partial (K'^{2})^{2}}|_{SP_{3}} = I_{5}^{'} =
-\frac{1}{12 \epsilon_{L}^{2}}[1 + \epsilon_{L}].
\end{equation}
The two-loop graph contributing to the 4-point function in the
isotropic situation is
\begin{eqnarray}
I_{4}\;\; =&& \int \frac{d^{m}k_{1}d^{m}k_{2}}
{(k_{1}^{2})^2 \bigl((K' - k_{1})^{2}\bigr)^2
\left(k_{2}^{2}\right) \bigl((k_{1} - k_{2} + k_{3}')^{2}\bigl)^2}\;\;,
\end{eqnarray}
where $K'= k_{1}' + k_{2}'$.
It can be integrated using the orthogonal approximation following
the same steps of the calculation of the anisotropic counterpart.
We simply quote the result
\begin{equation}
I_{4}(K'^{2}) = \frac{1}{2\epsilon_{L}^{2}}
[1 -\frac{\epsilon_{L}}{4}(1 + 2 L(K'^{2}))].
\end{equation}
At the symmetry point it is given by:
\begin{equation}
I_{4}(K'^{2}) = \frac{1}{2\epsilon_{L}^{2}}
[1 +\frac{3 \epsilon_{L}}{4}].
\end{equation}

It is worth to point out that the integrals $I_{3}$ and $I_{5}$ have not the
same leading singularities as in the usual $\phi^{4}$. Therefore, any attempt
to use the counterterms of the usual $\phi^{4}$ theory would lead to erroneous
results for the critical exponents in the isotropic case.

A recent calculation of the critical exponents $\eta_{L4}$ and
$\nu_{L4}$ for the isotropic case was presented by Diehl and Shpot
who fixed all the leading singularities equal to those appearing in the
standard $\phi^{4}$ theory \cite{Diehl4}. Moreover they ``choose''
the following angular factor:
\begin{equation}
F_{d}= 2^{1-d} \pi^{-\frac{d}{2}} \frac{\Gamma(5-\frac{d}{2})
\Gamma^{2}(\frac{d}{2} -2)}{\Gamma(d-4)}.\nonumber
\end{equation}
If one sets $d=8-\epsilon_{L}$, whenever the Feynman integral under
consideration presents a double pole in $\epsilon_{L}$, this angular
factor will give contribution to the simple poles in $\epsilon_{L}$.
This happens for example in the two-loop integrals $I_{2}^{2}$ and
$I_{4}$. Then, their calculation of these critical exponents cannot be
trusted, since further $\epsilon_{L}^{-1}$ terms were not taken into
account in the evaluation of those integrals. It seems that this factor
was used to reproduce the original critical exponents $\eta_{L4}$ and
$\nu_{L4}$ from the seminal paper \cite{Ho-Lu-Sh}.

Here the geometric angular factor is determined simply by requiring that
only $I_{2}$ has the same leading singularity as in the standard $\phi^{4}$
theory. In that case, it is simply given by the area of the $m$-dimensional
sphere.

We can now discuss the fixed points and critical exponents for
arbitrary $m$-axial behavior. Note that in the isotropic case,
only $I_{2}$ and $I_{4}$ have the same leading singularities as
in the standard $\phi^{4}$ field theory. This will lead to a
different fixed point for the isotropic behavior, as it is going
to be shown in a moment.

\section{The critical exponents for the anisotropic behaviors}

\subsection{The dissipative approximation}
The critical exponents were first calculated using this
approximation for the uniaxial case and the generalization was
soon presented for the $m$-axial case. Details can be found in
Refs. \cite{AL1,AL2}. Here we shall simply quote the results.

The fixed point at two-loop order is given by:
\begin{equation}
u^{\ast}=\frac{6}{8 + N}\,\epsilon_L\Biggl\{1 + \epsilon_L
\,\Biggl[ \Biggl(\frac{4(5N + 22)}{(8 + N)^{2}} - 1\Biggr )\,[i_{2}]_m -
\frac{(2 + N)}{(8 + N)^{2}}\Biggr]\Biggr\}\;\;.
\end{equation}
It can be used to obtain the critical exponents
$\eta_{L2}$ and $\nu_{L2}$:

\begin{eqnarray}
&&\eta_{L2}= \frac{1}{2}\epsilon_L^2\,\frac{2 + N}{( 8 + N)^2}\nonumber\\
&&\nonumber\\
&&\qquad +\;\, \epsilon_L^3\,
\frac{(2+N)}{(8 + N)^{2}}\,\Biggl[
\Biggl(\frac{4(5N + 22)}{(8 + N)^{2}} - \frac{1}{2}\Biggr )\,[i_{2}]_m
+\frac{1}{8-m}-\frac{2+N}{(8+N)^2}\Biggr]\;\;,
\end{eqnarray}

\begin{eqnarray}
&&\nu_{L2} =\frac{1}{2} + \frac{1}{4}\epsilon_L\,
\frac{ 2 + N }{8 + N}\nonumber\\
&&\nonumber\\
&&\; + \,\frac{1}{8}\frac{(2 + N)} {( 8 + N)^3}
\,\Biggl[2 (14N+40)\,[i_2]_m-2(2+N)+(8+N)(3+N)\Biggr]
\epsilon_L^2\;.
\end{eqnarray}

Using the scaling law
$\gamma_{L} = \nu_{L2}(2 - \eta_{L2})$, the exponent $\gamma_L$ is

\begin{eqnarray}
&&\gamma_{L} =
1 + \frac{1}{2}\epsilon_L\,\frac{2 + N}{8 + N}\nonumber\\
&&\nonumber\\
&&\qquad  +\;\; \frac{1}{4}\frac{(2 + N)}{(8 + N)^{3}}\,
\Biggl[12+8N+N^2+4\,[i_2]_m\,(20+7N)\Biggr]\,
\epsilon_L^2\,.
\end{eqnarray}

The evaluation of Feynman diagrams used to obtain these results have some
inconveniences as discussed before: the introduction of diagram factors for
the integrals $I'_{3}$ and $I'_{5}$ (which are divergent when $m=8$) is the
main trouble.

The RG analysis by Mergulh\~ao and Carneiro has been used
in a attempt to extend the calculation for all $m$. Using the 
fact that the quartic momenta scale including $\sigma$ and the
quadratic external momenta scales are equal Diehl and Shpot considered
the anisotropic problem for general $m$ \cite{Diehl1,Diehl2}. 
In their first work \cite{Diehl1}, they worked directly in
position space. After that, using a hybrid approach, going to 
coordinate or momentum space using the free propagator (scaling
function) in coordinate space to make the transition according to 
the necessity, they calculated the
critical exponents using a minimal subtraction procedure which sets
the external quartic momenta scales to zero. However, there is a small
discrepancy among their results for the critical exponents
in the cases $m=2,6$ when compared with Mergulh\~ao and Carneiro's results
using normalization conditions \cite{Diehl2}. For the anisotropic
cases $m \neq d$, the exponents are given in terms of integrals to be
performed numerically. These numerical integrals are meaningful solely
if one separates the integration limits on the variable 
${\bf v}=\sigma_{0}{\bf x}_{\parallel} {\bf x}_{\perp}$ using
the scaling and related functions in the coordinate space representation in the
integrand up to the maximum value of $|{\bf v}|$ at $|v_{0}|=9.3$, and
replacing the asymptotic value of these functions for greater values
of ${\bf v}$. Note that as the quartic and quadratic external
momenta are not independent, $\sigma$ {\it cannot} be taken
dimensionless as done by these authors following the invalid argument
by Mergulh\~ao and Carneiro. Thus, they erroneously concluded that the
isotropic case could be encompassed in their expressions for the
critical exponents in the limit $d \rightarrow m$ close to 8.

Furthermore, this alternative semianalytical method
has some drawbacks. First, setting the quadratic momenta scale to
zero makes impossible the transition from the anisotropic to the isotropic
case, since the quadratic momenta scale is absent in this case and
renormalization group transformations are defined only through the variation
of the quartic momenta scale. Second, unfortunately the
expression of the critical exponents for the anisotropic case are rather
cumbersome, given in terms of integrals to be performed numerically. Clearly
the most convenient answer should present analytical coefficients for each
order in $\epsilon_{L}$, in analogy to the usual standard $\phi^{4}$ theory
describing the Ising model.

The calculation of the critical exponents using the dissipative
approximation presented here has been criticized by Diehl and
Shpot \cite{Diehl3} because of the constraint introduced
in the quartic loop momenta in higher loop Feynman integrals.
Despite of this criticism, this approximation is in
good agreement with recent high-precision numerical data based in Monte
Carlo simulations for the ANNNI model \cite{Pleim,Le1}.

\subsection{The orthogonal approximation}

We turn now our attention to the most general approximation
for calculating all the critical exponents. In order to check
the results, we must calculate the critical exponents using
two different renormalization schemes, namely the normalization
conditions and minimal subtraction of dimensional poles. It is
going to be shown now that the critical indices are the same
irrespective of the use of either renormalization prescription.

\subsubsection{Normalization conditions and critical exponents}

We defined the bare coupling constants and renormalization functions as
\begin{mathletters}
\begin{eqnarray}
&& u_{o \tau} = u_{\tau} (1 + a_{1 \tau} u_{\tau} + a_{2 \tau} u_{\tau}^{2}) ,\\
&& Z_{\phi (\tau)} = 1 + b_{2 \tau} u_{\tau}^{2} + b_{3 \tau} u_{\tau}^{3} ,\\
&& \bar{Z}_{\phi^{2} (\tau)} = 1 + c_{1 \tau} u_{\tau} + c_{2 \tau} u_{\tau}^{2} ,
\end{eqnarray}
\end{mathletters}
where the constants $a_{i \tau}, b_{i \tau}, c_{i \tau}$ depend on Feynman
integrals calculated at the convenient symmetry points. Depending on
the symmetry point, we can calculate critical exponents corresponding to
correlations perpendicular or parallel to the competing $m$-dimensional
subspace.

The beta-functions and renormalization constants can be rewritten in
terms of the constants defined above in the following form:
\begin{mathletters}
\begin{eqnarray}
&& \beta_{\tau}  =  -\tau \epsilon_{L}u_{\tau}[1 - a_{1 \tau} u_{\tau}
+2(a_{1 \tau}^{2} -a_{2 \tau}) u_{\tau}^{2}],\\
&& \gamma_{\phi (\tau)} = -\tau \epsilon_{L}u_{\tau}[2b_{2 \tau} u_{\tau}
+ (3 b_{3 \tau}  - 2 b_{2 \tau} a_{1 \tau}) u_{\tau}^{2}],\\
&& \bar{\gamma}_{\phi^{2} (\tau)} = \tau \epsilon_{L}u_{\tau}[c_{1 \tau}
+ (2 c_{2 \tau}  - c_{1 \tau}^{2} - a_{1 \tau} c_{1 \tau})u_{\tau}    ].
\end{eqnarray}
\end{mathletters}

It is easy to obtain the coefficients above as functions of
the integrals calculated at the symmetry points. They are given by
\begin{mathletters}
\begin{eqnarray}
&& a_{1 \tau} = \frac{N+8}{6 \epsilon_{L}}[1 + [i_{2}]_{m} \epsilon_{L}] ,\\
&& a_{2 \tau} = (\frac{N+8}{6 \epsilon_{L}})^{2}
+ [\frac{(N+8)^{2}}{18}[i_{2}]_{m} - \frac{(3N+14)}{24}]
\frac{1}{\epsilon_{L}} ,\\
&& b_{2 \tau} = -\frac{(N+2)}{144 \epsilon_{L}}[1 +
(2 [i_{2}]_{m} + \frac{1}{4}) \epsilon_{L}], \\
&& b_{3 \tau} = -\frac{(N+2)(N+8)}{1296 \epsilon_{L}^{2}} +
\frac{(N+2)(N+8)}{108 \epsilon_{L}}(-\frac{1}{4} [i_{2}]_{m} +
\frac{1}{48}), \\
&& c_{1 \tau} = \frac{(N+2)}{6 \epsilon_{L}}[1 + [i_{2}]_{m} \epsilon_{L}], \\
&& c_{2 \tau} = \frac{(N+2)(N+5)}{36 \epsilon_{L}^{2}}
+ \frac{(N+2)}{3 \epsilon_{L}}[\frac{(N+5)}{3} [i_{2}]_{m} - \frac{1}{4}].
\end{eqnarray}
\end{mathletters}
This is enough to obtain the fixed points at $O(\epsilon_{L}^{2})$.
They are defined by $\beta_{\tau}(u_{\tau}^{*}) = 0$. All the
integrals calculated at the symmetry points $SP_{1}$ and $SP_{2}$
look the same. The factor of $\tau=1,2$ drops out at the fixed points,
implying that the renormalization group transformations performed
either over $\kappa_{1}$ or $\kappa_{2}$ will flow to the same fixed
point given by ($u_{1}^{*} = u_{2}^{*} \equiv u^{*}$)
\begin{equation}
u^{\ast}=\frac{6}{8 + N}\,\epsilon_L\Biggl\{1 + \epsilon_L
\,\Biggl[ - [i_{2}]_m + \frac{(9N + 42)}{(8 + N)^{2}}\Biggr]\Biggr\}\;\;.
\end{equation}
The surprising feature of this approximation is that the critical
exponents do not depend on $[i_{2}]_m$, as the ones obtained using the
dissipative approximation. The functions $\gamma_{\phi (1)}$ and
$\bar{\gamma}_{\phi^{2} (1)}$ can be written as
\begin{eqnarray}
&& \gamma_{\phi (1)} = \frac{(N+2)}{72} [1 + (2 [i_{2}]_{m} + \frac{1}{4})
\epsilon_{L}]u_{1}^{2} - \frac{(N+2)(N+8)}{864} u_{1}^{3}, \\
&& \bar{\gamma}_{\phi^{2} (1)} = \frac{(N+2)}{6} u_{1}[1
+ [i_{2}]_{m} \epsilon_{L} - \frac{1}{2} u_{1}].
\end{eqnarray}
Replacing the value of the fixed point inside these equations, using the
relation among these functions and the critical exponents $\eta_{L2}$ and
$\nu_{L2}$, we find:
\begin{eqnarray}
&& \eta_{L2}= \frac{1}{2} \epsilon_{L}^{2}\,\frac{N + 2}{(N+8)^2}
[1 + \epsilon_{L}(\frac{6(3N + 14)}{(N + 8)^{2}} - \frac{1}{4})] ,\\
&& \nu_{L2} =\frac{1}{2} + \frac{(N + 2)}{4(N + 8)} \epsilon_{L}
+  \frac{1}{8}\frac{(N + 2)(N^{2} + 23N + 60)} {(N + 8)^3} \epsilon_{L}^{2}.
\end{eqnarray}
Notice that the coefficient of each power of $\epsilon_{L}$ is the same
as in the pure $\phi^{4}$ describing the Ising-like behavior. The
reduction to the $m=0$ case is even simpler using this approximation
than the reduction to the $m=0$ case using the dissipative approximation.
Since the functions $\gamma_{\phi (2)}= 2 \gamma_{\phi (1)}$ and
$\bar{\gamma}_{\phi^{2} (2)} = 2 \bar{\gamma}_{\phi^{2} (1)}$ as a
consequence of $\beta_{2} = 2 \beta_{1}$, we immediately conclude that
\begin{eqnarray}
&& \eta_{L4}= \epsilon_{L}^{2}\,\frac{(N + 2)}{(N+8)^2}
[1 + \epsilon_{L}(\frac{6(3N + 14)}{(N + 8)^{2}} - \frac{1}{4})] ,\\
&& \nu_{L4} =\frac{1}{4} + \frac{(N + 2)}{8(N + 8)} \epsilon_{L}
+  \frac{1}{16}\frac{(N + 2)(N^{2} + 23N + 60)} {(N + 8)^3} \epsilon_{L}^{2}.
\end{eqnarray}
Thus, at $O(\epsilon_{L}^{3})$, the relation $\eta_{L4} = 2 \eta_{L2}$
is valid. At  $O(\epsilon_{L}^{2})$,  the relation
$\nu_{L4} = \frac{1}{2} \nu_{L2}$ is fulfilled. Thus the strong anisotropic
scale invariance \cite{He} is {\it exact} to the perturbative order
considered here. The other
exponents can be read from the scaling relations. As discussed before,
they are ($\alpha_{L2} = \alpha_{L4} = \alpha_{L}$, etc.):
\begin{eqnarray}
&& \gamma_{L} = 1 + \frac{(N+2)}{2(N + 8)} \epsilon_{L}
+\frac{(N + 2)(N^{2} + 22N + 52)}{4(N + 8)^{3}} \epsilon_{L}^{2}, \\
&& \alpha_{L} = \frac{(4 - N)}{2(N + 8)} \epsilon_{L}
- \frac{(N + 2)(N^{2} + 30N + 56)}{4(N + 8)^{3}} \epsilon_{L}^{2} ,\\
&& \beta_{L} = \frac{1}{2} - \frac{3}{2(N + 8)} \epsilon_{L}
+ \frac{(N + 2)(2N + 1)}{2(N + 8)^{3}} \epsilon_{L}^{2} ,\\
&& \delta_{L} = 3 + \epsilon_{L}
+ \frac{(N^{2} + 14N + 60)}{2(N + 8)^{2}} \epsilon_{L}^{2}.
\end{eqnarray}

Note that all these exponents reduce to the Ising-like case when $m=0$. In
order to check the correctness of these exponents, it is convenient to
calculate them in another renormalization procedure, as we shall see next.

\subsubsection{Minimal subtraction and critical exponents}

Usually, in the minimal subtraction renormalization scale, one can have
more than one coupling, but just one momenta scale, called in most textbooks
$\mu$\cite{Si} and named $\kappa$ here. The dimensional redefinition
performed for the quartic external momenta, allows the description of
the anisotropic case with two independent momenta scales. The coupling
constant has two independent flows, induced by $\kappa_{1}$ and $\kappa_{2}$.

If we want to calculate the critical exponents along the competition axes,
we set the quadratic external momenta perpendicular to the competing
subspace to zero. Thus, one can introduce the quartic momenta scale
$\kappa_{2}$ in order to compute the normalization functions for arbitrary
quartic external momenta and demanding that the dimensional poles
(logarithmic divergences in the momenta) be minimally subtracted.
On the other hand, the calculation of critical exponents perpendicular
to the competing axes can be performed by setting the quartic external
momenta to zero, introducing $\kappa_{1}$, calculating the
normalization functions for arbitrary quadratic external momenta and
requiring minimal subtraction.

In this section, we are not going to calculate explicitly the critical
exponents. Instead, we are going to calculate the fixed point as well as
the functions $\gamma_{\phi (\tau)}$ and  $\bar{\gamma}_{\phi^{2} (\tau)}$
at the fixed point. As these functions at the fixed point are
universal, they should be equal to the ones obtained using
normalization conditions, leading to the same exponents in either
renormalization scheme.

The dimensionless bare couplings and the renormalization functions
are defined in minimal subtraction by

\begin{mathletters}
\begin{eqnarray}
&& u_{0 \tau} = u_{\tau}[1 + \sum_{i=1}^{\infty} a_{i \tau}(\epsilon_{L})
u_{\tau}^{i}], \\
&& Z_{\phi (\tau)} = 1 + \sum_{i=1}^{\infty} b_{i \tau}(\epsilon_{L})
u_{\tau}^{i},\\
&& \bar{Z}_{\phi^{2} (\tau)} = 1 + \sum_{i=1}^{\infty} c_{i \tau}(\epsilon_{L})
u_{\tau}^{i}.
\end{eqnarray}
\end{mathletters}
The renormalized vertices
\begin{mathletters}
\begin{eqnarray}
&& \Gamma_{R (\tau)}^{(2)}(k_{\tau}, u_{\tau}, \kappa_{\tau}) = Z_{\phi (\tau)}
\Gamma_{(\tau)}^{(2)}(k_{\tau}, u_{0 \tau}, \kappa_{\tau}), \\
&& \Gamma_{R (\tau)}^{(4)}(k_{i \tau}, u_{\tau}, \kappa_{\tau}) = Z_{\phi (\tau)}^{2} \Gamma_{(\tau)}^{(4)}(k_{i \tau}, u_{0 \tau}, \kappa_{\tau}), \\
&& \Gamma_{R (\tau)}^{(2,1)}(k_{1 \tau}, k_{2 \tau}, p_{\tau};
u_{\tau}, \kappa_{\tau}) = \bar{Z}_{\phi^{2} (\tau)}
\Gamma_{(\tau)}^{(2,1)}(k_{1 \tau}, k_{2 \tau}, p_{\tau}, u_{0 \tau}, \kappa_{\tau}),
\end{eqnarray}
\end{mathletters}
are finite when $\epsilon_{L} \rightarrow 0$, order by order in
$u_{\tau}$. Note that the external momenta into the bare vertices
are mutiplied by $\kappa_{\tau}^{-1}$. Recall that $k_{i1} = p_{i}$
are the external momenta perpendicular to the competing axes, whereas
$k_{i2} = k'_{i}$ are the external momenta parallel to the
$m$-dimensional subspace. The coefficients $a_{i \tau}(\epsilon_{L}),
b_{i \tau}(\epsilon_{L})$ and $c_{i \tau}(\epsilon_{L})$ are obtained
by requiring that the poles in $\epsilon_{L}$ be minimally subtracted.
The bare vertices can now be expressed as

\begin{mathletters}
\begin{eqnarray}
&& \Gamma_{(\tau)}^{(2)}(k_{\tau}, u_{0 \tau}, \kappa_{\tau}) =
k_{\tau}^{2 \tau}(1- B_{2 \tau} u_{0 \tau}^{2} + B_{3 \tau}u_{0 \tau}^{3}), \\
&& \Gamma_{(\tau)}^{(4)}(k_{i \tau}, u_{0 \tau}, \kappa_{\tau}) =
\kappa_{\tau}^{\tau \epsilon} u_{0 \tau}
[1- A_{2 \tau} u_{0 \tau}
+ (A_{2 \tau}^{(1)} + A_{2 \tau}^{(2)})u_{0 \tau}^{2}], \\
&& \Gamma_{(\tau)}^{(2,1)}(k_{1 \tau}, k_{2 \tau}, p_{\tau};
u_{0 \tau}, \kappa_{\tau}) = 1 - C_{1 \tau} u_{0 \tau}
+ (C_{2 \tau}^{(1)} + C_{2 \tau}^{(2)}) u_{0 \tau}^{2}.
\end{eqnarray}
\end{mathletters}
Notice that $B_{2 \tau}$ is proportional to the integral $I_{3}$ and
$B_{3 \tau}$ is proportional to $I_{5}$. Note that if $\tau =1$, all
the integrals should be replaced by their values at zero quartic external
momenta. In case $\tau=2$, those integrals are calculated at zero quadratic
external momenta.

Explicitly, the coefficients are given by the following integrals:

\begin{mathletters}
\begin{eqnarray}
&& A_{1 \tau} = \frac{(N+8)}{18}[ I_{2}(\frac{k_{1 \tau} + k_{2 \tau}}
{\kappa_{\tau}}) +  I_{2}(\frac{k_{1 \tau} + k_{3 \tau}}
{\kappa_{\tau}}) + I_{2}(\frac{k_{2 \tau} + k_{3 \tau}}
{\kappa_{\tau}})] ,\\
&& A_{2 \tau}^{(1)} = \frac{(N^{2} + 6N + 20)}{108}
[I_{2}^{2}(\frac{k_{1 \tau} + k_{2 \tau}}{\kappa_{\tau}})
+  I_{2}^{2}(\frac{k_{1 \tau} + k_{3 \tau}}
{\kappa_{\tau}}) + I_{2}^{2}(\frac{k_{2 \tau} + k_{3 \tau}}
{\kappa_{\tau}})] ,\\
&& A_{2 \tau}^{(2)} = \frac{(5N + 22)}{54}
[I_{4}( \frac{k_{i \tau}}{\kappa_{\tau}}) + 5   \;permutations] ,\\
&& B_{2 \tau} = \frac{(N+2)}{18}I_{3}(\frac{k_{\tau}}{\kappa_{\tau}}) ,\\
&& B_{3 \tau} =
\frac{(N+2)(N+8)}{108}I_{5}(\frac{k_{\tau}}{\kappa_{\tau}}) ,\\
&& C_{1 \tau} = \frac{N+2}{18}[ I_{2}(\frac{k_{1 \tau} + k_{2 \tau}}
{\kappa_{\tau}}) +  I_{2}(\frac{k_{1 \tau} + k_{3 \tau}}
{\kappa_{\tau}}) + I_{2}(\frac{k_{2 \tau} + k_{3 \tau}}
{\kappa_{\tau}})] ,\\
&& C_{2 \tau}^{(1)} = \frac{(N+2)^{2}}{108}
[I_{2}^{2}(\frac{k_{1 \tau} + k_{2 \tau}}{\kappa_{\tau}})
+  I_{2}^{2}(\frac{k_{1 \tau} + k_{3 \tau}}
{\kappa_{\tau}}) + I_{2}^{2}(\frac{k_{2 \tau} + k_{3 \tau}}
{\kappa_{\tau}})] ,\\
&& C_{2 \tau}^{(2)} = \frac{N+2}{36}[I_{4}(\frac{k_{i
\tau}}{\kappa_{\tau}}) + 5   \;permutations].
\end{eqnarray}
\end{mathletters}

This is sufficient to determine the normalization constants to the
loop order desired. Requiring minimal subtraction of dimensional poles
for the renormalized vertex parts quoted above, all the logarithmic
integrals in the external momenta appearing in $I_{2}, I_{3}, I_{4}$,
and $I_{5}$ cancell each other. The result is that the normalization
functions and coupling constants can be expressed in the form:

\begin{mathletters}
\begin{eqnarray}
&& u_{0 \tau} = u_{\tau}(1 + \frac{(N+8)}{6 \epsilon_{L}} u_{\tau}
+ [\frac{(N+8)^{2}}{36 \epsilon_{L}^{2}} - \frac{(3N+14)}{24
\epsilon_{L}}] u_{\tau}^{2}), \\
&& Z_{\phi (\tau)} = 1 - \frac{N+2}{144 \epsilon_L} u_{\tau}^{2}
+ [-\frac{(N+2)(N+8)}{1296  \epsilon_{L}^{2}} + \frac{(N+2)(N+8)}{5184
\epsilon_{L}}] u_{\tau}^{3}, \\
&& \bar{Z}_{\phi^{2} (\tau)} = 1 + \frac{N+2}{6 \epsilon_L} u_{\tau}
+ [\frac{(N+2)(N+5)}{36 \epsilon_{L}^{2}} - \frac{(N+2)}{24
\epsilon_{L}}] u_{\tau}^{2}).
\end{eqnarray}
\end{mathletters}

From the renormalization functions one can obtain:
\begin{eqnarray}
&& \gamma_{\phi (\tau)} = \tau [\frac{(N+2)}{72}u_{\tau}^{2}
- \frac{(N+2)(N+8)}{1728}u_{\tau}^{3}],\\
&& \bar{\gamma}_{\phi^{2} (\tau)} = \tau \frac{(N+2)}{6} u_{\tau}
[ 1 - \frac{1}{2} u_{\tau}].
\end{eqnarray}

The fixed points are defined by $\beta_{\tau}(u_{\tau}^{*}) =
0$. Then,  it is found that the fixed points generated by
renormalization group transformations over either $\kappa_{1}$
or $\kappa_{2}$ are the same and is given by:
\begin{equation}
u_{\tau}^{\ast}=\frac{6}{8 + N}\,\epsilon_L\Biggl\{1 + \epsilon_L
\,\Biggl[\frac{(9N + 42)}{(8 + N)^{2}}\Biggr]\Biggr\}\;\;.
\end{equation}

Substitution of this result into the renormalization constants
will give at the fixed point $\gamma_{\phi (\tau)}^{*}= \eta_{\tau}$,
where $\eta_{\tau}$ are given by Eqs. (184) and (186). In addition,
we have
\begin{equation}
\bar{\gamma^{*}}_{\phi^{2} (\tau)} = \tau \frac{(N+2)}{(N+8)} \epsilon_{L}
[ 1 + \frac{6(N+3)}{(N+8)^{2}} \epsilon_{L}].
\end{equation}
This leads to the same exponents $\nu_{\tau}$ given in Eqs. (185) and
(187), obtained there via normalization conditions. Therefore, we have
proven the consistency of this picture for the anisotropic Lifshitz
critical behavior, since the critical indices are independent of the
renormalization procedure.

\subsubsection{Discussion}

The exponent $\eta_{L2}$ obtained
here agrees with the calculation performed independently by Mukamel
\cite{Muka}. Nevertheless, the exponent $\eta_{L4}$ presented here is at
variance with Mukamel's \cite{Muka} and, therefore, with the result
obtained by Hornreich and Bruce \cite{Ho-Bru} since both works agree
with each other.

We are now in position to compare our results with those obtained for the
ANNNI model in three-dimensional space ($\epsilon_{L}=1.5$) representing
the uniaxial ($m=1$) case using Monte Carlo simulations \cite{Pleim}.
From the numerical viewpoint, there is no sensitive difference among the
results presented either using the dissipative approximation or the orthogonal
approximation for the critical exponents perpendicular to the competition
axes. Within the two significative algarisms precision the exponents using
either approximation are given by $\eta_{L2}=0.04$ and $\nu_{L2} = 0.73$.

The deviations start in the calculation of $\gamma_{L2}= \gamma_{L}$. In the
dissipative approximation the $\epsilon_{L}$-expansion yields
$\gamma_{L}=1.45$. A numerical interpretation has been proposed recently
in order to improve the results obtained via the $\epsilon_{L}$-expansion
when the perturbative parameter $\epsilon_{L}$ is greater than 1 \cite{Le1}.
There it was argued that the neglected
$O(\epsilon_{L}^{3})$ could be relevant to the calculation of, say,
$\gamma_{L}$. The basic idea is to replace the numerical
values of $\nu_{L2}$ and $\eta_{L2}$ directly into the scaling laws in order
to obtain the other critical exponents. In this way one obtains
$\gamma_{L}=1.43$, $\alpha_{L} = 0.18$ and $\beta_{L} = 0.20$.

On the other hand, using the $\epsilon_{L}$-expansion results for
$\gamma_{L}$, $\alpha_{L}$ and $\beta_{L}$ obtained via the orthogonal
approximation one finds $\gamma_{L} = 1.42$, $\alpha_{L} = 0.05$
and $\beta_{L} = 0.26$. The numerical ``ansatz''described above gives
again $\gamma_{L}=1.43$, $\alpha_{L} = 0.18$ and $\beta_{L} = 0.20$,
since $\eta_{L2}$ and $\nu_{L2}$ have the same numerical values in either
approximation.  These numbers should be compared with the newest Monte
Carlo simulations output, namely, $\gamma_{L} = 1.36 \pm 0.03$,
$\alpha_{L}=0.18 \pm 0.02$ and $\beta_{L} = 0.238 \pm 0.005$.

Thus, the greater the mean field values for the exponents, the better
are the their numerical values using the $\epsilon_{L}$-expansion when
$\epsilon_{L}$ is not a small number. Otherwise, the numerical ``ansatz''
yields a rather good agreement with the Monte Carlo results, as in the case
for the exponents $\alpha_{L}$ and $\beta_{L}$. This shows that the new
results displayed here are consistent with the best numerical values
available for the ANNNI model.

\section{The critical exponents for the isotropic systems}

As the isotropic behavior presents just one external momenta scale, its
analysis is simpler than the one used to describe the anisotropic behavior,
where two external momenta scales are present. Besides, the only manner to
attack this problem is to use the orthogonal approximation, for the
dissipative approximation does not work as it was discussed before.

\subsection{Critical exponents in normalization conditions}

The bare coupling constants and renormalization functions are defined as
\begin{mathletters}
\begin{eqnarray}
&& u_{0 3} = u_{3} (1 + a_{1 3} u_{3} + a_{2 3} u_{3}^{2}) ,\\
&& Z_{\phi (3)} = 1 + b_{2 3} u_{3}^{2} + b_{3 3} u_{3}^{3} ,\\
&& \bar{Z}_{\phi^{2} (3)} = 1 + c_{1 3} u_{3} + c_{2 3} u_{3}^{2} ,
\end{eqnarray}
\end{mathletters}
where the constants $a_{i 3}, b_{i 3}, c_{i 3}$ depend on Feynman
integrals calculated at the symmetry point named henceafter $SP_{3}$.
Only the external momenta scale $\kappa_{3}$ parallel to the
competing $m$-dimensional subspace arises in this isotropic case.

The beta-function and renormalization constants are written in
terms of the constants defined above in the following manner:
\begin{mathletters}
\begin{eqnarray}
&& \beta_{3}  =  - \epsilon_{L}u_{3}[1 - a_{1 3} u_{3}
+2(a_{1 3}^{2} -a_{2 3}) u_{3}^{2}],\\
&& \gamma_{\phi (3)} = - \epsilon_{L}u_{3}[2b_{2 3} u_{3}
+ (3 b_{3 3}  - 2 b_{2 3} a_{1 3}) u_{3}^{2}],\\
&& \bar{\gamma}_{\phi^{2} (3)} = \epsilon_{L}u_{3}[c_{1 3}
+ (2 c_{2 3}  - c_{1 3}^{2} - a_{1 3} c_{1 3})u_{3}    ].
\end{eqnarray}
\end{mathletters}

The coefficients above are obtained as functions of
the integrals calculated at the symmetry point. They read
\begin{mathletters}
\begin{eqnarray}
&& a_{1 3} = \frac{N+8}{6 \epsilon_{L}}[1 + \frac{1}{4} \epsilon_{L}] ,\\
&& a_{2 3} = (\frac{N+8}{6 \epsilon_{L}})^{2}
+ [\frac{2N^{2} + 23N + 86}{144 \epsilon_{L}}] ,\\
&& b_{2 3} = -\frac{(N+2)}{288 \epsilon_{L}}[1 + \frac{5}{8} \epsilon_{L}], \\
&& b_{3 3} = -\frac{(N+2)(N+8)}{2592 \epsilon_{L}^{2}} -
\frac{(N+2)(N+8)}{20736 \epsilon_{L}}, \\
&& c_{1 3} = \frac{(N+2)}{6 \epsilon_{L}}[1 + \frac{1}{4} \epsilon_{L}], \\
&& c_{2 3} = \frac{(N+2)(N+5)}{36 \epsilon_{L}^{2}}
+ \frac{(N+2)(2N+7)}{144 \epsilon_{L}}.
\end{eqnarray}
\end{mathletters}
The fixed point is defined by $\beta_{3}(u_{3}^{*}) = 0$. Therefore, it is
given by
\begin{equation}
u_{3}^{\ast}=\frac{6}{8 + N}\,\epsilon_L\Biggl\{1 + \epsilon_L
\frac{1}{2}\Biggl[ - \frac{1}{2} + \frac{(9N + 42)}{(8 + N)^{2}}\Biggr]\Biggr\}\;\;.
\end{equation}
Note that this fixed point is different from that appearing in the
anisotropic behavior and cannot be obtained from it in a smooth way.
The functions $\gamma_{\phi (3)}$ and
$\bar{\gamma}_{\phi^{2} (3)}$ can be written as
\begin{eqnarray}
&& \gamma_{\phi (3)} = \frac{(N+2)}{144} [1 + \frac{5}{8}
\epsilon_{L}]u_{3}^{2} - \frac{(N+2)(N+8)}{3456} u_{3}^{3}, \\
&& \bar{\gamma}_{\phi^{2} (3)} = \frac{(N+2)}{6} u_{3}[1
+ \frac{1}{4} \epsilon_{L} - \frac{1}{4} u_{3}].
\end{eqnarray}
Replacing the value of the fixed point inside these equations, using the
relation among these functions and the critical exponents $\eta_{L4}$ and
$\nu_{L4}$, we find:
\begin{eqnarray}
&& \eta_{L4}= \frac{1}{4} \epsilon_{L}^{2}\,\frac{N + 2}{(N+8)^2}
[1 + \epsilon_{L}(\frac{3(3N + 14)}{(N + 8)^{2}} - \frac{1}{8})] ,\\
&& \nu_{L4} =\frac{1}{4} + \frac{(N + 2)}{16(N + 8)} \epsilon_{L}
+  \frac{1}{256}\frac{(N + 2)(N^{2} + 23N + 60)} {(N + 8)^3} \epsilon_{L}^{2}.
\end{eqnarray}
These exponents are different from those originally obtained in
Ref.\cite{Ho-Lu-Sh}. The coefficient of the $\epsilon_{L}^{2}$ term in the
exponent $\eta_{L4}$ is positive, consistent with its counterpart in the
anisotropic cases as well as in the Ising-like case. One learns that
only the quartic momenta is not sufficient to induce its change of sign. The
exponent $\nu_{L4}$ agrees at $O(\epsilon_{L})$ with that presented in
Ref.\cite{Ho-Lu-Sh} but naturally disagrees at $O(\epsilon_{L}^{2})$,
since it depends on the value of $\eta_{L4}$ at $O(\epsilon_{L}^{2})$.
Besides, the critical index $\eta_{L4}$ is obtained at $O(\epsilon_{L}^{3})$
here for the first time.

Now using the scaling relations derived for the isotropic case we obtain
immediately

\begin{eqnarray}
&& \gamma_{L4} = 1 + \frac{(N+2)}{4(N + 8)} \epsilon_{L}
+\frac{(N + 2)(N^{2} + 19N + 28)}{64(N + 8)^{3}} \epsilon_{L}^{2}, \\
&& \alpha_{L4} = \frac{(4 - N)}{4(N + 8)} \epsilon_{L}
+ \frac{(N + 2)(N^{2} + 9N + 68)}{32(N + 8)^{3}} \epsilon_{L}^{2} ,\\
&& \beta_{L4} = \frac{1}{2} - \frac{3}{4(N + 8)} \epsilon_{L}
- \frac{(N + 2)(N^{2} + N + 108)}{64(N + 8)^{3}} \epsilon_{L}^{2} ,\\
&& \delta_{L4} = 3 + \frac{1}{2} \epsilon_{L}
+ \frac{(N^{2} + 14N + 60)}{8(N + 8)^{2}} \epsilon_{L}^{2}.
\end{eqnarray}
These exponents are obtained here for the first time at
$O(\epsilon_{L}^{2})$. Formerly the lack of a set of scaling laws for the
isotropic case did not allow these findings. In order to check these
results, let us analyse the situation using the minimal subtraction
scheme.

\subsection{Critical exponents in minimal subtraction}

We proceed analogously as in the isotropic case. We just replace the
subscript $\tau =3$ and keep in mind that the Feynman integrals are
calculated in the isotropic case $d=m$ close to 8. Minimal subtraction
of dimensional poles in the renormalized vertex $\Gamma_{R (3)}^{(4)}$
implies that the bare dimensionless coupling constant can be expressed
in the form:
\begin{equation}
u_{0 3} = u_{3}[1 + \frac{(N+8)}{6 \epsilon_{L}} u_{3} +
(\frac{(N+8)^{2}}{36 \epsilon_{L}^{2}} - \frac{(3N+14)}{48 \epsilon_{L}})
u_{3}^{2}].
\end{equation}

The fixed point can be easily found to be
\begin{equation}
u_{3}^{*} = \frac{6}{(N+8)} \epsilon_{L} +
\frac{9(3N+14)}{(N+8)^{3}} \epsilon_{L}^{2}.
\end{equation}

The normalization constants are given by:
\begin{eqnarray}
&& Z_{\phi (3)} = 1 - \frac{(N+2)}{288 \epsilon_{L}} u_{3}^{2} \nonumber\\
&& + [-\frac{(N+2)(N+8)}{2592 \epsilon_{L}^{2}} + \frac{(N+2)(N+8)}{20736
\epsilon_{L}}] u_{3}^{3},\\
&& \bar{Z}_{\phi^{2} (3)} = 1 + \frac{(N+2)}{6 \epsilon_{L}} u_{3} \nonumber\\
&& + [\frac{(N+2)(N+5)}{36 \epsilon_{L}^{2}} - \frac{(N+2)}{48 \epsilon_{L}}]
u_{3}^{2}.
\end{eqnarray}

The functions $\gamma_{\phi (3)}$ and $\bar{\gamma}_{\phi^{2} (3)}$ are
given by the following expressions:

\begin{mathletters}
\begin{eqnarray}
&& \gamma_{\phi (3)} = \frac{(N+2)}{144} u_{3}^{2} - \frac{(N+2)(N+8)}{6912} u_{3}^{3},\\
&& \gamma_{\phi^{2} (3)} = \frac{(N+2)}{6}(u_{3} - \frac{1}{4} u_{3}^{2}).
\end{eqnarray}
\end{mathletters}

Using these results the function $\gamma_{\phi (3)}^{*}$ at the fixed point
yields the value of $\eta_{L4}$ as obtained in (207), whereas the function
$\bar{\gamma}_{\phi^{2} (3)}^{*}$ at the fixed point reads
\begin{equation}
\bar{\gamma}_{\phi^{2} (3)}^{*} = \frac{(N+2)}{(N+8)} \epsilon_{L} [1 +
\frac{3(N+3)}{(N+8)^{2}} \epsilon_{L}],
\end{equation}
which is the same as that obtained in the fixed point using normalization
conditions and leads to the same critical exponent $\nu_{L4}$ from (208)
as the reader is invited to check. Therefore, the complete
equivalence between the two renormalization schemes is assured.

Notice that the critical exponent $\eta_{L4}$ for the isotropic case is
different from the original result Ref.\cite{Ho-Lu-Sh}. Since we have
checked our results using two distinct renormalization schemes as shown
above, the critical indices presented by those authors should be checked
using more than one renormalization procedure in order to clarify
this discrepancy.

\section{Conclusions and perspectives}

All the critical exponents for the $m$-axial Lifshitz critical behavior
for the anisotropic ($1\leq m \leq d-1$) and the isotropic
($d=m$ close to 8) cases are explicitly derived at $O(\epsilon_{L}^{2})$.
We have shown that up to the loop order considered in this work strong
anisotropic scaling theory holds since the relations
$\nu_{L4} = \frac{1}{2}\nu_{L2}$ and $\eta_{L4} = 2 \eta_{L2}$ are exact.
The exponents associated to critical correlations perpendicular to the
competing axes easily reduces to the Ising-like exponents when $m=0$, the
only difference being the perturbation parameter $\epsilon_{L}$ replacing the
usual $\epsilon$ in noncompeting systems. These relations imply that the
new scaling laws, obtained here using two independent renormalization group
transformations, reduce to the ones previously found by Hornreich, Luban and
Shtrikman \cite{Ho-Lu-Sh}.

Moreover, to our knowledge this is the first time
that all the exponents for the isotropic behavior are obtained explicitly
through the use of the new scaling relations presented \cite{Le1}. Besides,
they are shown explicitly not to be recoverable from the anisotropic
situation in the limit $d \rightarrow m$. The structure of the Feynman
integrals in the isotropic case indicates that it deserves a especial
treatment when compared with the anisotropic situation as clarified in this
article.

The new results for the calculation of arbitrary loop Feynman integrals are
obtained by demanding that they are homogeneous functions of arbitrary
external momenta. Even though the calculations are carried out in a given
order in perturbation theory, the author is convinced, however, that the
conclusions hold to all orders.

The simple analytical expressions for each
coefficient in the $\epsilon_{L}$-expansion of the critical indices are
rather encouraging to proceed the evaluation of other universal amounts,
like critical amplitudes \cite{Le2}. It would be interesting to compare
some experimental results available for MnP like the specific heat
critical amplitude ratio \cite{Be} with theoretical calculations within
the context of an $\epsilon_{L}$-expansion using the techniques described
in the present work. In addition, a thorough RG analysis to prove that all
amplitude ratios are indeed universal for the Lifshitz critical behavior
was not done yet. Actually, the idea presented in this work might be
suitable to demonstrate the universal character of the above mentioned
critical ratios and calculate all of them.

Other problems can be pursued using the present method. The treatment
of finite-size effects for the Lifshitz behavior can
be devised in analogy to the noncompeting situation
\cite{Bre-Zinn, Nemi}. The systems may be finite (or semi-infinite)
along one (or several) of their dimensions, but they are of infinite
extent in the remaining directions. Examples include systems which are
finite in all directions, such as a (hyper) cube of size $L$, and systems
which are of infinite size in $d'=d-1$ dimensions but are either of
finite thickness $L$ along the remaining direction ($d$-dimensional layered
geometry) or of a semi-infinite extension. The presence of geometrical
restrictions on the domain of systems also requires the introduction of
boundary conditions (periodic, antiperiodic, Dirichlet and Neumann)
satisfied by the order parameter on the surfaces. In particular, the
validity limits of the $\epsilon_{L}$-expansion for these systems and the
approach to bulk criticality in a layered geometry can be studied \cite{Le3}.

Recently, typical surface phenomena in noncompeting systems were generalized
to competing systems using Monte Carlo simulations for the ANNNI model
\cite{Pleim2}. However, as far as the Lifshitz behavior is concerned, a
theoretical description of these systems is still lacking. The
field-theoretical framework just presented might be useful to address
this problem.

The quest towards a generalization of the Lifshitz universality classes
whenever arbitrary momenta powers arise in the Lagrangian (1) as the effect
of further competition is quite a fascinating issue \cite{Sel}. It is
expected that it can be solved along the same lines described in this
work \cite{Le4}.

In summary, we have described the Lifshitz critical behavior in its
complete generality in what concerns its critical exponents.
We have presented new field theory renormalization group
methods which resulted in new analytical expressions for all the critical
indices in the isotropic as well as in the anisotropic cases at least at
$O(\epsilon_{L}^{2})$. We hope our findings will be useful to unveil further
issues related to the physics of competing systems.

\section{Acknowledgments}

The author acknowledges financial support from FAPESP, grant number
00/06572-6 and Departamento de F\'\i sica Matem\'atica da Universidade de
S\~ao Paulo for the use of its facilities. He also thanks
L. C. de Albuquerque and N. Berkovits for discussions.

\end{document}